\documentclass[12pt,showpacs,preprintnumbers,superscriptaddress,nofootinbib]{revtex4}
\usepackage{amssymb}
\usepackage{amsmath, graphicx}
\usepackage{dcolumn}
\usepackage{bm}
\usepackage{epstopdf}
\usepackage{color}
\usepackage{xcolor}

\newcommand{\hs}{\hspace*{0.3cm}}

\newcommand{\be}{\begin{equation}}
\newcommand{\ee}{\end{equation}}
\newcommand{\bea}{\begin{eqnarray}}
\newcommand{\eea}{\end{eqnarray}}
\newcommand{\ben}{\begin{enumerate}}
\newcommand{\een}{\end{enumerate}}
\newcommand{\bit}{\begin{itemize}}
\newcommand{\eit}{\end{itemize}}
\newcommand{\bde}{\begin{widetext}}
\newcommand{\ede}{\end{widetext}}
\newcommand{\nn}{\nonumber}
\newcommand{\crn}{\nonumber \\}
\newcommand{\eq}{\eqref}

\newcommand{\al}{\alpha}
\newcommand{\la}{\lambda}
\newcommand{\bet}{\beta}
\newcommand{\ga}{\gamma}
\newcommand{\va}{\varphi}

\newcommand{\fr}{\frac}

\newcommand{\bc}{\begin{center}}
\newcommand{\ec}{\end{center}}
\newcommand{\Ga}{\Gamma}
\newcommand{\de}{\delta}
\newcommand{\De}{\Delta}

\newcommand{\si}{\sigma}

\newcommand{\crb}[1]{{\color{blue} #1 }}

\usepackage{soul}
\usepackage{cancel}

\input{tcilatex}

\begin{document}

\title{
Constraining   heavy neutral gauge boson $Z'$  in  the  3 - 3 - 1 models
 by  weak charge data of Cesium and proton}
\author{H. N. Long}
\email{hoangngoclong@tdtu.edu.vn}
\affiliation{Theoretical Particle Physics and Cosmology Research Group, Advanced Institute
	of Materials Science,  Ton Duc Thang University, Ho Chi Minh City 700000, Vietnam}
\affiliation{Faculty of Applied Sciences,
	Ton Duc Thang University, Ho Chi Minh City 700000, Vietnam}

\author{N. V. Hop}
\email{nvhop@ctu.edu.vn}
\affiliation{Department of  Physics, Can Tho University, Can Tho 900000, Vietnam}
\affiliation{Faculty of Physics, Hanoi Pedagogical University 2, Phuc Yen, Vinh Phuc 280000, Vietnam}

\author{L.T. Hue\footnote{Corresponding author}}\email{lthue@iop.vast.vn}
\affiliation{Institute for Research and Development, Duy Tan University, Da Nang City 550000, Vietnam}
\affiliation{Institute of Physics,   Vietnam Academy of Science and Technology, 10 Dao Tan, Ba Dinh, Hanoi 100000, Vietnam}

\author{N.~T.~T.~ Van}\email{thuvan@assoc.iop.vast.ac.vn}
\affiliation{Institute of Physics,   Vietnam Academy of Science and Technology, 10 Dao Tan, Ba Dinh, Hanoi 100000, Vietnam}
\date{\today }

\begin{abstract}
The  recent experimental data of the weak charges  of Cesium and proton
is  analyzed
in the framework of the models based on the  $\mbox{SU}(3)_C\times \mbox{SU}(3)_L \times
\mbox{U}(1)_X$ (3-3-1) gauge group,  including the 3-3-1 model with CKS mechanism (3-3-1CKS) and the general 3-3-1 models with	arbitrary  $\beta$  (3-3-1$\beta$) with three Higgs triplets. We will show that at the TeV  scale, the mixing among neutral gauge bosons plays
significant  effect.
Within the present  values of the weak charges of Cesium and proton we get
the  lowest  mass bound  of the  extra heavy neutral gauge boson   to be 1.27 TeV.
 The results derived  from the weak charge data, perturbative limit of Yukawa coupling of the top quark, and the relevant Landau poles    favor  the models with $\beta = \pm \fr{1}{\sqrt{3}}$ and $\beta = 0$ while ruling out the ones with  $\beta= \pm \sqrt{3}$. In addition, there are some hints showing that in the 3-3-1 models, the third quark family should be treated differently  from the first twos.

\end{abstract}
	
\maketitle

\allowdisplaybreaks

\section{Introduction}\label{intro}

Nowadays, the experimental data on neutrino masses and mixing as well as on Dark Matter (DM) lead to fact that the Standard Model (SM) must be extended.
 Among the beyond SM extensions, the models based on the
$\mbox{SU}(3)_C\times \mbox{SU}(3)_L \times \mbox{U}(1)_X$ gauge
group \cite{Valle:1983dk,Pisano:1991ee,Frampton:1992wt,Foot:1992rh,Foot:1994ym,Hoang:1995vq,Hoang:1996gi}
(3 - 3 - 1 models)
are attractive  in the following senses.   First of all,  these models are  concerned with the
search of an explanation for the number of fermion generations to be three, when  the QCD asymptotic freedom is combined.
Some other advantages of the 3-3-1 models are: i)  the electric
charge quantization is solved \cite{deSousaPires:1998jc,VanDong:2005ux}, ii) there
are several sources of CP violation \cite{Montero:1998yw,Montero:2005yb}%
, and iii) the
strong-CP problem is solved due to the natural Peccei-Quinn symmetry \cite{Pal:1994ba,Dias:2002gg,Dias:2003zt,Dias:2003iq}.

There are two main versions of
the 3 - 3 - 1 models which depend on the parameter $\bet$ in the electric charge operator
\be
Q=T_3+\beta T_8+X\, . \label{qoperator}
\ee
If $\bet = \sqrt{3}$, this is the minimal version \cite{Pisano:1991ee,Frampton:1992wt,Foot:1992rh}, and
$\bet = - \fr 1{ \sqrt{3}}$
corresponds to the 3-3-1 model with right-handed neutrinos
\cite{Valle:1983dk,Foot:1994ym,Hoang:1995vq,Hoang:1996gi}.

At present, we
still face an old problem of explanation of
hierarchies and structure of the fermion sector.
However, in the above models,  most researches on the 3-3-1 models
are not concerned with vast different masses among the generations  (see
references in Ref.\cite{CarcamoHernandez:2017cwi}).
It is well known that the Yukawa interactions are not enough for producing fermion masses and mixings.
According to our best of knowledge, the first work for  solving the mentioned puzzles in quark sector is
in Ref.~\cite{Froggatt:1978nt} named Froggatt-Nielsen mechanism. Recently, the new mechanism based on
 sequential loop suppression mechanism, is more natural since its
suppression factor is arisen from loop factor $l\approx (1/4\pi)^2 $. The above mentioned mechanism is called by CKS
- the names of its authors \cite{CarcamoHernandez:2016pdu}.
The Froggatt-Nielsen mechanism was implemented to the 3-3-1 model in Ref.~\cite{Huitu:2017ukq}.
 In recent work Ref. ~\cite{CarcamoHernandez:2017cwi}
the CKS mechanism has been implemented to the 3-3-1 model with $\bet = - \fr 1{ \sqrt{3}}$, and it is
 interesting to note that the derived
model is renormalizable.  We name it the  3-3-1\ CKS  model for short.  In the Ref. \cite{Long:2018dun},
the Higgs and gauge sectors of the model are explored.
From the experimental data on the $\rho$ parameter, the bound on the scale of the first step of the spontaneous symmetry breaking (SSB) in the 3-3-1 CKS is in the range
of 6 TeV \cite{Long:2018dun}.
There also exist helpful  relations among masses of gauge bosons, this is essential point for
  the model phenomenology.

At present, the new neutral gauge boson $Z'$ is a very attractive subject in Particle Physics due to  potential discovery
 of right-handed neutrinos through its mediation \cite{Freitas:2018vnt}.  Within its mass  around 2.5 TeV, the simulation
 shows that it may  be discovered at the LHC.   Hence it is necessary to study
 more deeply different aspects  to fix the mass as well as
    properties  of $Z'$. To fix the model parameters, one often looks at well known observables such as the $\rho$ parameter, mass differences of neutral mesons,
   and deviation of weak charge of nucleus, etc. So, in this paper we focus on  the latter subject.

Recently, new  constraints of the $Z'$ mass around  4 TeV  have been reported from  studying the  $Z'$ decays into the SM lepton pairs,  based on the new LHC Run 2 data  \cite{Aaboud:2017buh,Aaboud:2017sjh,Sirunyan:2018exx, Aad:2019fac} \footnote{ We thank the referee for reminding us this point.}.   On the other hand,  a recent study on a particular 3-3-1 model  argued that the lower bounds of $Z'$ mass can be significantly  smaller than those obtained from LHC, if  other decay channels of $Z'$ into new particles  are included  \cite{Coriano:2018coq}. We will follow this particular framework, i.e.  the new constraints of $Z'$ will be omitted in our discussion. A more general  dependence of  the lower bounds of $Z'$ mass in 3-3-1 models on the LHC data will be studied in the future.

  The parity violation in weak interactions
 was known for long time ago. In the SM, it can be seen from the atomic parity violation (APV) caused by
the neutral gauge boson $Z$.  In the beyond Standard Model (BSM), the APV gets additional contribution
from new heavy neutral  gauge bosons $Z^\prime$.
Therefore, the data on APV, especially of the Cesium  ($^{133}_{55}Cs $)  being stable atom,  is an effective channel for probing
the new neutral gauge boson $Z^\prime$.  This is our aim in this work.

The experimental data on    the APV  in  Cesium atom~\cite{Bennett:1999pd} has
caused extensive interest and reviews~\cite{Rosner:2001ck, Ginges:2003qt,Bouchiat:2004sp, Guena:2005uj,Davoudiasl:2012qa, Erler:2014fqa}.
Parity violation in the SM results from
exchanges of weak gauge bosons, namely,  in electron-hadron
neutral-current processes. The parity violation is due to
the vector axial-vector interaction in the effective Lagrangian.
The measurement is stated in terms of the weak charge
$Q_W$, which parameterizes the parity violating Lagrangian.
Due to the extra neutral gauge bosons, in the BSM, the  weak charge of an isotope (X)  gets additional
value which is called by deviation defined as follows
\be
\De Q_W(^A_ZX) \equiv Q_W^{ \mathrm{BSM}}(^A_ZX) - Q_W^{\mathrm{SM}}(^A_ZX)\, .
\label{eq181}
\ee
For the concrete stable isotope Cesium (Cs), it is reported  recently from experiment as \cite{Dzuba:2012kx, Tanabashi:2018oca}
\be
\label{first}
Q^{\mathrm{exp}}_W ( ^{133}_{55} {\rm Cs} ) = -72.62 \pm 0.43.
\ee

 Comparing  to the SM
prediction $Q_W^{\mathrm{SM}}(^{133}_{55}Cs)=-73.23\pm 0.01$ \cite{Erler:2013xha,Tanabashi:2018oca} yields  the deviation $\De Q_W$ as follows  \cite{Dzuba:2012kx}
\be
\De Q_W(^{133}_{55}Cs) \equiv Q^{\mathrm{exp}}_W(^{133}_{55}{\rm Cs}) - Q_W^{\rm SM}(^{133}_{55}{\rm Cs})
= 0.61 \pm 0.43\, ,
\label{eqo51}
\ee
which is $1.4~ \si$ away from the SM prediction. This value
has been widely used for analysis of possible new physics, where it is assumed that the BSM can be explained the experimental value of the weak charge  $Q_W(^{133}_{55}Cs)$.

On the other hand, the weak charge  of an atom is formulated as a function of  the two independent contributions of light quarks $u$ and $d$,  the experimental weak charge values of the two distinguishable  isotopes  will result   in  different allowed regions of the parameter space defined by a BSM.   Hence, combining result of allowed regions from  experimental weak charge data   of  Cesium and proton will be more strict than the previous one.  Recently, the experiments of   parity-violation in electron scattering (PVES), see a review in~\cite{Souder:2015mlu},  have  determined  the latest value of the proton's  weak charge, namely   $Q^{\mathrm{exp}}_W(^1_1p)=0.0719\pm 0.0045$ \cite{Androic:2018kni}.  It was shown to be in great agreement with the SM prediction, $Q^{\mathrm{SM}}_W(^1_1p)=0.0708\pm 0.0003$.  The deviation from the SM is
\be
\De Q_W(^{1}_{1}p)
= 0.0011 \pm 0.0045.
\label{eq_DQWp}
\ee
 Considering a BSM
  containing an additional heavy neutral gauge
 boson $Z'$ apart from the SM one $Z$,  a theoretical  deviation of $Q_W$ from the SM prediction
 for an isotope $^A_Z X$  is given by
\bea \label{eq_DeQBSM}
 \De Q^{\mathrm{BSM}}_W(^A_ZX)  & \simeq &
\left[ 2Z - A  +  4  Z \left( \fr{s_W^4}{1- 2 s^2_W}\right)\right] \De \rho
 \crn
&+ &4s_\phi \left\{(A+Z) \left[g_A(e) g'_V(u) +g'_A(e) g_V(u)\right] \right.
\label{del2}\\
&+&\left. (2A - Z) \left[g_A(e) g'_V(d) +g'_A(e) g_V(d)\right]  \right\}
\crn&-& 4\left(\fr{M_{Z_1}^2}{M^2_{Z_2}}\right)[(A+Z)g'_A(e)g'_V(u) +(2A - Z) g'_A(e)g'_V(d) ],\nn
\eea
where
$s_\phi \equiv \sin\phi$ corresponds to the $Z-Z'$ mixing of the SM and  new heavy neutral gauge bosons $Z$
 and $Z'$ that create the two physical states  $Z_{1,2}$ with masses $M_{Z_{1,2}}$.

 Notations in Eq.~\eq{eq_DeQBSM}
are based on the vector-axial (V-A) currents of neutral
  gauge  bosons defined by the well-known
 Lagrangian
 \bea \label{eqVff}
 \mathcal{L}_{Vff}& = & \fr{g}{2c_W}\sum_{f} \overline{f}\ga^\mu(g_V(f) -\ga_5 g_A(f)f Z_\mu \crn
 &+&\fr{g}{2c_W}\sum_{f} \overline{f}\ga^\mu(g'_V(f) -\ga_5 g'_A(f)f Z'_\mu,
 \eea
where the summation is taken over the fermions of the BSM,   $g = e/ s_W$ is the $SU(2)_L$ gauge coupling of the SM.

The formula \eq{eq_DeQBSM} has been checked in details  by us (see appendix \ref{deviation}) based on original
calculation in Ref.~\cite{Altarelli:1991ci} that concerned for $U(1)$ gauge extensions of the SM.
However,  it is also valid for other non-Abelian
 gauge extensions including 3-3-1 models \cite{Hoang:2000jy,CarcamoHernandez:2005ka,Gutierrez:2005rq,Dong:2006cn,Salazar:2007ym,Gauld:2013qja,Buras:2013dea,Martinez:2014lta}.
 Especially,  the formulas for arbitrary $\beta$ given in Ref.~\cite{CarcamoHernandez:2005ka} was corrected in Ref.~\cite{Martinez:2014lta}
 following a recent correction of $Z-Z'$ mixing angle \cite{Buras:2014yna}.   Using the same notations our formula \eq{eq_DeQBSM} contains two factors 4 instead of 16 in the expression of the weak charge used in Ref.~\cite{Martinez:2014lta}.
 Additionally,  the numerical investigation in Ref.~\cite{Martinez:2014lta} used the old experimental data of the Cs weak charge \cite{Beringer:1900zz},  which is very well consistent with the SM prediction.  On the other hand,  the  new constraint given in  Eq.~\eq{first} is significantly different from the previous \cite{Beringer:1900zz}, and   implies a certain deviation from the SM.
 Therefore, a new investigation based on the latest experimental data of both weak charges of Cesium and proton will result in new information of allowed regions of the parameter spaces in the 3-3-1 models.

Taking into account the SM gauge couplings
\be
 g_A(e) = -\fr 1 2 \, , \hs  g_V(u)=\fr 1 2 -\fr{4 s_W^2}{3}  \, , \hs  g_V(d)=-\fr 1 2 +\fr{2s_W^2} 3 \, ;
 \label{t3}
 \ee
the experimental value of the Weinberg angle at the $M_Z$ scale \cite{Tanabashi:2018oca}  $s_W^2= 0.23122 $,
$\left( \fr{s_W^4}{1- 2 s^2_W}\right) =0.0994544 $;
and  the scale dependence of the gauge couplings $g$  in Eq. \eq{eqVff}, the expression~\eq{eq_DeQBSM} is written
 in the more general form
\bea \De Q^{\mathrm{BSM}}_W(^A_ZX)  & \simeq &
 -  \left(A- 2.39782 \times Z  \right) \De \rho
 \crn
&- &2s_\phi \left\{A \left[  2 g'_V(d) +
g'_V(u) + g'_A(e) \right] \right.
\label{del4}\\
&-&\left.  Z \left[g'_A(e)\times 1.07512 +  g'_V(d) -  g'_V(u)\right]  \right\} \times \frac{g(M_{Z_2})}{g(M_{Z_1})}
\crn&-& 4g'_A(e)\left(\fr{M_{Z_1}^2}{M^2_{Z_2}}\right)\left\{A\left[2 g'_V(d)
+g'_V(u)\right] + Z\left[ g'_V(u) -g'_V(d)\right] \right\} \times \frac{g^2(M_{Z_2})}{g^2(M_{Z_1})}
\,  ,
\nn
\eea
where $g(M_{Z_{1,2}})$  are respective gauge couplings of the $Z_{1,2}$ at their mass scales. We emphasize that Eq.~\eq{del4} contains
 major improvements from the original version \cite{Altarelli:1991ci}, see detailed discussion in appendix \ref{deviation}.
The above formula is also applicable for
the models based on  $\mbox{SU}(3)_C\times \mbox{SU}(3)_L \times \mbox{U}(1)_X$ gauge
group, where effect of scale dependence was mentioned but the $Z-Z'$ mixing was ignored~\cite{Gauld:2013qja,Buras:2013dea}.  The subject was also considered earlier in Refs. \cite{Hoang:2000jy,Dong:2006cn}, but for only the minimal and economical 3-3-1 versions, respectively. The formula
 \eq{del4} is different from those  used to investigate APV in 3-3-1 models in  Refs.~\cite{Martinez:2014lta}, where the scale dependence
  of neutral gauge couplings are also  taken into account.  Furthermore, in the light of   new experimental results  of  weak charges  and rho parameter~\cite{Tanabashi:2018oca}, the parameter spaces of the 3-3-1 models will be  re-investigated. Instead of Ref.~\cite{Martinez:2014lta}, where  only model C introduced in Ref.~\cite{Martinez:2006gb} was paid attention using the APV of $Q_W(Cs)$,  we will discuss all  allowed regions of the three parameter spaces corresponding  to the three models A, B, and C, based on the latest experimental data of both $Q_W(Cs)$ and $Q_W(p)$. The effects of the perturbative limit of  top quark Yukawa coupling on the parameter space will also be included.  The combination resulting from the three mentioned ingredients will affect differently the parameter spaces of the three 3-3-1 models A,B,C.  Hence, it may suggest which models can be survived or ruled out, instead of the common acceptance in literature  that prefers the model  A,  where the heavy quark family containing the top quark is treated differently from the two lighter ones.

 The further plan of this paper is as follows. Sect.~\ref{mcks} is devoted to the 3-3-1 CKS model where the particle content
 is introduced. In this section, the gauge boson masses and mixing  are also discussed, and the couplings
  between neutral gauge bosons $Z$ and $Z'$ and fermions are presented. In Sect. \ref{cks331}, we consider the deviation
  of weak charge for Cesium in the 3-3-1 CKS, from which the lower bound on the $M_{Z_2}$ is derived.  Sect.
  \ref{331beta} is devoted for the model  3-3-1$\beta$~\cite{Ochoa:2005ih,CarcamoHernandez:2005ka}.  In this section, we will focus on different kinds  of quark assignments listed in Ref.~\cite{CarcamoHernandez:2005ka},   where the heavy flavor quarks $t$ and $b$ behave
   differently from other ones (representation A) or
  the light quarks $u$ and $d$ do the same (representation C).
  The analytic expressions of the deviations $\De Q^{\mathrm{BSM}}_W(^A_ZX) $
predicted by  the  models will be combined with the latest  data of APV and PVES to investigate  allowed regions of the parameter spaces, which can result in the possibility of  surviving or ruling out the model under consideration. We make a conclusion in the last
  section - section \ref{conclusion}. Two appendices show in detailed steps how to derive the analytic expressions of the  weak charges  in the general case and the particular case of the 3-3-1$\beta$ model.

\section{Atomic parity violation in the
3 - 3 - 1 CKS model}
\label{mcks}
	
In this  section the needed ingredients for investigating the  weak charges  predicted by the 3-3-1 CKS model are discussed.

\subsection{Particle content}

As in the ordinary 3-3-1 model without exotic electric charges, the quark
sector contains two quark generations transforming as  antitriplet and one remaining generation transforming as triplet
under  $SU(3)_L$ subgroup. The other extra quarks transform as singlet under above mentioned subgroup. The quantum numbers
of the quark sector are summarized in Table \ref{bangquark}.

\begin{table}[tbp]
\caption{Quark assignments under
$SU(3)_L, U(1)_X, U(1)_{L_g},  Z_4, Z_2$
 and the values of generalized lepton number $L_{g}$ (all quarks are in triplets under $SU(3)_C $)
}
\resizebox{16cm}{!}{
\renewcommand{\arraystretch}{1.2}
\begin{tabular}{|c|c|c|c|c|c|c|c|c|c|c|c|c|c|c|c|c|c|c|}
\hline
 & $Q_{1L}$ & $Q_{2L}$ & $Q_{3L}$ & $U_{1R}$ & $U_{2R}$ & $U_{3R}$ & $T_R$ & $D_{1R}$ & $D_{2R}$ & $D_{3R}$ & $J_{1R}$ & $J_{2R}$ & $\widetilde{T}_{1L}$ & $\widetilde{T}_{1R}$ & $\widetilde{T}_{2L}$ & $\widetilde{T}_{2R}$ & $B_L$ & $B_R$ \\ \hline
 $SU(3)_L$&$3^*$&$3^*$&$3$&1&1&1&1&1&1&1&1&1&1&1&1&1&1&1
 \\ \hline
 $X$&0&0&$\fr 1 3$ &$\fr 2 3$&$\fr 2 3$&$\fr 2 3$&$\fr  2 3$&$-\fr 1 3$&$-\fr 1 3$&$-\fr 1 3$&$-\fr 1 3$&$-\fr 1 3$&$\fr  2 3$&$\fr  2 3$&$\fr  2 3$&$\fr  2 3$&$-\fr 1 3$&$-\fr 1 3$\\
 \hline
$L_g$ & $\fr 2 3 $ & $\fr 2 3 $ & $-\fr 2 3 $ & $0$ & $0$ & $0$ & $-2$ & $0$ & $0$ & $0$ & $2$ & $2$ & $0$ & $0$ & $0$ & $0$ & $0$ & $0$ \\ \hline
 $Z_4$ & $-1$ & $-1$ & $1$ & $1$ & $-i$ & $1$ & $1$ & $1$ & $1$ & $1$ & $-1$ & $-1$ & $i$ & $1$ & $i$ & $1$ & $-1$ & $-1$ \\ \hline
  $Z_2$ & $1$ & $1$ & $1$ & $1$ & $1$ & $-1$ & $-1$ & $1$ & $1$ & $1$ & $-1$ & $-1$ & $1$ & $1$ & $1$ & $1$ & $1$ & $1$ \\ \hline
\end{tabular}}
\label{bangquark}
\end{table}
As seen from Table \ref{bangquark},  in the model under consideration,  all extra quarks have  electric charges  of quarks in the SM. As shown in Ref.~\cite{CarcamoHernandez:2017cwi}, the spontaneous symmetry breaking (SSB) provides masses for only extra quarks as well as top quark.
The remaining quarks get masses
by radiative corrections. To explain why top quark gets mass at the tree level but bottom quark does not get, the
reason lies in the behaviour of their right-handed components under the symmetry $Z_2$: $U_{3R}$  is odd, while $D_{3R}$  is even. It is
crucial for the forbiddance of unwanted terms.

The content of the leptonic sector is summarized in Table \ref{banglepton}. As in the quark sector, the extra leptons:
$E_i, i = 1,2,3$, $N_i, i = 1,2,3$ and $\Psi_R$ get masses at the tree level. Table  \ref{banglepton} also shows that under the $Z_2$, right-handed
components of the charged leptons in the second (muon) and the third (tauon) generations are even, while for the first generation, it is odd. That is why
tauon and muon get masses at the one-loop level, but the electron gets mass at two-loop correction \cite{CarcamoHernandez:2017cwi}. Table \ref{banglepton}
also shows that the extra neutral leptons $N_i, i = 1,2,3$ have lepton number \emph{opposite} to those of  ordinary leptons.

\begin{table}[tbp]
\caption{Lepton assignments under
$SU(3)_L, U(1)_X, U(1)_{L_g},  Z_4, Z_2$
 and the values of generalized lepton number $L_{g}$  (all leptons are   singlets under $SU(3)_C $)
}
\resizebox{16cm}{!}{
\renewcommand{\arraystretch}{1.2}
\begin{tabular}{|c|c|c|c|c|c|c|c|c|c|c|c|c|c|c|c|c|}
\hline
 & $L_{1L}$ & $L_{2L}$ & $L_{3L}$ & $e_{1R}$ & $e_{2R}$ & $e_{3R}$ & $E_{1L}$ & $E_{2L}$ & $E_{3L}$ & $E_{1R}$ & $E_{2R}$ & $E_{3R}$ & $N_{1R}$ & $N_{2R}$ & $N_{3R}$ & $\Psi_R$ \\ \hline
 $SU(3)_L$&3&3&3&1&1&1&1&1&1&1&1&1&1&1&1&1\\ \hline
 $X$ &$-\fr 1 3$&$-\fr 1 3$&$-\fr 1 3$&-1&-1&-1&-1&-1&-1&-1&-1&-1&0&0&0&0 \\ \hline
 $L_g$ & $\fr 1 3 $ & $\fr 1 3 $ & $\fr 1 3 $ & $1$ & $1$ & $1$ & $1$ & $1$ & $1$ & $1$ & $1$ & $1$ & $-1$ & $-1$ & $-1$ & $1$ \\ \hline
 $Z_4$ & $i$ & $i$ & $i$ & $-i$ & $-i$ & $-i$ & $1$ & $i$ & $i$ & $-i$ & $-i$ & $-i$ & $i$ & $i$ & $i$ & $1$ \\ \hline
  $Z_2$ & $-1$ & $1$ & $1$ & $-1$ & $1$ & $1$ & $-1$ & $1$ & $1$ & $-1$ & $1$ & $1$ & $-1$ & $-1$ & $-1$ & $-1$ \\ \hline
\end{tabular}}
\label{banglepton}
\end{table}

The Higgs sector contains three scalar triplets $\chi $, $\eta $ and $\rho $
and seven singlets $\va _1^0$,$\ \va _2 ^0$, $\xi^0$,$\
\phi _1^+$,$\ \phi _2 ^+$, $\phi _3^+$ and $\phi _4^+$.
The content of the Higgs sector is presented in Table \ref{bangscalar}.

\begin{table}[tbp]
\caption{Scalar assignments under
$SU(3)_L, U(1)_X, U(1)_{L_g},  Z_4, Z_2$
 and the values of generalized lepton number $L_{g}$.
}
\begin{tabular}{|c|c|c|c|c|c|c|c|c|c|c|}
\hline
 & $\chi$ & $\eta$ & $\rho$ & $\va_1^0$ & $\va_2^0$ & $\phi_1^+$ & $\phi_2^+$ & $\phi_3^+$ & $\phi_4^+$ & $\xi^0$\\ \hline
 $SU(3)_L$&3&3&3&1&1&1&1&1&1&1\\ \hline
  $X$ &$-\fr 1 3$&$-\fr 1 3$&$\fr 2 3$&0&0&1&1&1&1&0
  \\ \hline
 $L_g$ & $\fr 4 3$ & $-\fr 2 3 $ & $-\fr 2 3 $ & $0$ & $0$ & $0$ & $-2$ & $-2$ & $-2$ & $-2$ \\ \hline
$Z_4$ & $1$ & $1$ & $-1$ & $-1$ & $i$ & $i$ & $-1$ & $-1$ & $1$ & $1$ \\ \hline
$Z_2$ & $-1$ & $-1$ & $1$ & $1$ & $1$ & $1$ & $1$ & $-1$ & $-1$ & $1$\\ \hline
\end{tabular}\vspace{-0.1cm}
\label{bangscalar}
\end{table}

We note  that, in contradiction  with
ordinary 3-3-1  model, the neutral component of the $\rho$ triplet does not have a vacuum expectation value (VEV).
 That is why  the charged leptons
do not get masses at the tree level. From Table \ref{bangscalar}, it follows that $\chi$ triplet has generalized lepton number $L_g$ \cite{CarcamoHernandez:2017cwi,Chang:2006aa} different from those of $\eta$ and $\rho$ triplets. This leads to the fact that the bottom
elements of the $\eta$ and $\rho$ triplets as well as two first rows of  the $\chi$ have lepton number equal to 2, the same as $\phi^+_i, i = 2,3,4$ and $\xi$ do.

To close this section, we remind that after SSB, the charged and non-Hermitian gauge bosons get masses as below ~\cite{Long:2018dun}
\be
m^2_W = \fr{g^2}{4}v_\eta^2 \, , \hs  M^2_{X^0} = \fr{g^2}{4}%
\left(v^2_\chi +v_\eta^2\right) \, , \hs  M^2_{Y} = \fr{g^2}{4}%
v_\chi^2\, ,  \label{eq185}
\ee
where we have used the following notations
\be
W_\mu^\pm = \fr{1}{\sqrt{2}}\left( A_{\mu 1} \mp i A_{\mu 2} \right)\, ,
\hs  Y_\mu^\pm = \fr{1}{\sqrt{2}}\left( A_{\mu 6} \pm i A_{\mu
7} \right)\, , \hs  X_\mu^0 = \fr{1}{\sqrt{2}}\left( A_{\mu 4}
- i A_{\mu 5} \right)\, .  \label{eq91}
\ee
From \eq{eq91}, the following consequences are in order
\bea  && v_\eta   =  v  = 246 \,   \textrm{GeV} \, , \label{eq92}\\
&& M^2_{X^0} - M^2_Y  =  m^2_W \, .  \label{eq186}
\eea

Note that the value $\De Q_W$ depends on couplings of neutral gauge bosons $Z$ and $Z'$ with light quark
$u$ and $d$. Hence, we turn to the neutral current sector of the model.

\subsection{Neutral currents}

Looking at Eq. \eq{eq135}, one recognizes that some couplings between fermions and neutral gauge bosons $Z, Z'$
enter to the discrepancies.  The needed
interactions between fermions and gauge bosons are followed from a piece
\be
L_{\mbox{fermion \& gauge boson}} \supset \sum_f i \overline{f} \ga^\mu D_\mu f\, .
\label{eq1819}
\ee
Here, the covariant derivative is defined by
\be
D_\mu = \partial_\mu - i g A_{\mu a} T_a - i g_X X T_9B_\mu \, ,  \label{eq182}
\ee
 where $g$ and $g_X$ are the gauge coupling constants of the $SU(3)_L$ and $U(1)_X$ groups, respectively. Here, $T_a$ ($a=1,2,..,9$)
 are the generators of the $SU(3)$ group with gauge bosons $A_{\mu a}$. Corresponding to the $SU(3)_L$ representations, namely
  triplet, antitriplet, or singlet of the fermion, $T_a=\fr 1 2 \la_a, -\fr 1 2 \la^T_a$, or $0$. Furthermore, we choose the $U(1)_X$ generator
  as $T_9 = 1/\sqrt{6}\text{ diag}
(1,1,1)$ for both triplet and antitriplet, while $T_9=1/\sqrt{6}$ for singlets.
For the convenience, one rewrites \eq{eq182} as follows
\be
D_\mu = \partial_\mu - i g P_\mu^{CC} - i g P_\mu^{NC} \, ,  \label{eq1820}
\ee
where
\be
P_\mu^{CC} = \sum_{a = 1,2,4,5,6,7}T_a A_{\mu a}\, ,  \label{eq1821}
\ee
 and $P_\mu^{NC}$ is determined from
diagonal generators, namely
\be
P_\mu^{NC} = \sum_{a = 3,8} T_a A_{\mu a} + t X T_9 B_\mu, \,  \hs t \equiv
\fr{g_X}{g} = \fr{3\sqrt{2} \sin \theta_W(M_{Z^\prime})}{%
\sqrt{3-4\sin^2 \theta_W(M_{Z^\prime})}} \, .\label{eq_PNC}
\ee

Since  atom cesium is only composed of light quarks, namely $u$ and $d$ quarks and electron, therefore,
we just need to deal with these fermions.
The coupling constants relevant for calculations of
APV in the cesium atom for the SM and the 3 - 3 - 1 CKS model
 are presented
in Table \ref{tab1}.
\begin{table}[htb]
\caption{ Vector and axial-vector coupling constants relevant for APV in the SM and  3 - 3 - 1 CKS model}
\newcommand{\m}{\hphantom{$-$}}
\newcommand{\cc}[1]{\multicolumn{1}{c}{#1}}
\renewcommand{\tabcolsep}{2pc} 
\renewcommand{\arraystretch}{1.2} 
\medskip
\centering
\begin{tabular}{|c|c|}
\hline
Standard Model &  3-3-1\ CKS model   \\
\hline
$ g_A(e) =-\fr 1 2  $ & $g'_A(e) =  + \fr{1}{2\sqrt{3-4s^2_W}}  $ \\
\hline
$ g_V(u)=\fr 1 2 -\fr{4 s_W^2}3  $ &
$ g'_V(u)=\fr{-3 +8s_W^2}{6\sqrt{3-4s_W^2}}$  \\
\hline
$ g_V(d)=-\fr 1 2 +\fr{2s_W^2}3 $ &  $g'_V(d)= \fr{ -3+2 s_W^2}{6\sqrt{3-4s_W^2}} $\\
\hline
\end{tabular}
\label{tab1}
\end{table}

In the limit $v_\chi \gg v_\eta$, the $Z-Z^\prime$ mixing angle is  \cite{Long:2018dun}
\be
\tan \phi \simeq \fr{(1-2 s_W^2)\sqrt{3-4s^2_W}}{4 c^4_W}\left( \fr{%
v_\eta^2}{v_\chi^2} \right) \, .  \label{eq1817}
\ee

\subsection{Deviation of  the weak charge expression in the 3-3-1 CKS model}
\label{cks331}

Let us note that one of the most important observables
is the $\rho$ parameter defined as
\be
\rho =\fr{m^2_W}{c_W^2 M^2_{Z_1}}\, ,  \label{eq261}
\ee
where $\rho=1$ for the SM.
Let us analyze the expression in \eq{del4} with $\De \rho\equiv \rho-1$ for a BSM. The $\De \rho$ is determined by
\be \De \rho \simeq \al T,
\label{t}\ee
where $\al $ is the fine structure constant and $T$ is one of the Peskin-Takeuchi parameters~\cite{Peskin:1990zt}.
 The latter is given by
\be
T=   T_{Z Z'} +
T_{oblique}\, ,
\label{e301}
\ee
where the contribution from $Z-Z^\prime$ mixing $ T_{Z Z'}$ is as follows
\be
   T_{Z Z'}   \simeq  \fr{\tan^2
\phi}{\al}\left( \fr{M^2_{Z_2}}{M^2_{Z_1}} -1 \right)\, .
\label{t2}
\ee
The $T_{oblique}$ being an  oblique correction,  is  model dependent.

Applying Eq. \eq{del4} for Cesium yields
 \bea
\De Q_W(^{133}_{55} {\rm Cs})&=&-1.12004 \times \De \rho \crn
 &&- s_\phi  \left[422 \,  g'_V(d) + 376 \, g'_V(u) + 147.737\,  g'_A(e)\right] \times \fr{g(M_{Z_2})}{g(M_{Z_{1}})} \crn
  &&- g'_A(e)\left[844 . g'_V(d) +752.  g'_V(u) \right] \left(\fr{M^2_{Z_1}}{M^2_{Z_2}}\right)\, \times \fr{g^2(M_{Z_2})}{g^2(M_{Z_{1}})}.\label{del7}
\eea

 Taking values $g'_A(e), g'_A(d)$, and $g'_A(u)$  from Table \ref{tab1}, we get an expression for  $\De Q_W(^{133}_{55} {\rm Cs})$ predicted
  by the 3-3-1 CKS model
 \bea
\De Q_W^{\mathrm{CKS}}(^{133}_{55} {\rm Cs})&=&-1.12004 \times  \al( T^{\mathrm{CKS}}_{Z Z'} +T^{\mathrm{CKS}}_{oblique})\crn
&&+ \left[ s_\phi \times 122.655 \times \fr{g(M_{Z_2})}{g(M_{Z_{1}})}+ 120.743\left(\fr{M^2_{Z_1}}{M^2_{Z_2}}\right)\times \fr{g^2(M_{Z_2})}{g^2(M_{Z_{1}})}\right] \, .\label{del8}
\eea

Looking at  Eq.\eq{del8}, we see that when $M^2_{Z_2} \rightarrow \infty$,  the value
$\De Q_W^{CKS}(^{133}_{55} {\rm Cs})$ can be negative. However, it is very tiny.
According to Ref.~\cite{CarcamoHernandez:2005ka}, in the minimal model, the first term $\propto - 0.01$, while in
 Ref.~\cite{Buras:2014yna}, the $T_{oblique}$ is neglected. Following recent experimental data of $\De \rho$, which is in order
  of $\mathcal{O}(10^{-4})$, we accept the assumption in Ref.~\cite{Buras:2014yna}.

The weak charge of the proton is determined as
\bea
	\De Q_W^{\mathrm{CKS}}(^{1}_{1}p)&=&1.140 \De \rho  + \left[0.437 \times \fr{g(M_{Z_2})}{g(M_{Z_{1}})}+ 0.777\times \fr{g^2(M_{Z_2})}{g^2(M_{Z_{1}})}\right] \left(\fr{M^2_{Z_1}}{M^2_{Z_2}}\right) . \label{deQWp}
	\eea
For the model under consideration, the oblique correction has  the same form given in Ref. \cite{Long:2018dun,Hoang:1999yv}.
Combining with Eq.~(\ref{eq186}), ones get \cite{Long:2018dun}
\bea
\De \rho_{CKS}  &\simeq & \tan^2 \phi \left( \fr{ M^2_{Z^\prime }}{m^2_{Z }} -
1 \right) + \fr{3\sqrt{2}G_F}{16\pi^2} \left[ 2 M_{Y^+}^2 + m^2_W - \fr{%
2M_{Y^+}^2 (M_{Y^+}^2 +m^2_W) }{m^2_W} \ln \fr{(M_{Y^+}^2 +m^2_W)}{%
M_{Y^+}^2} \right]  \crn
& & - \fr{\al(m_Z)}{4\pi\ s^2_W} \left[\ t^2_W \ln \fr{(M_{Y^+}^2
+m^2_W)}{M_{Y^+}^2} + \fr{m_W^4}{2(M_{Y^+}^2 +m^2_W)^2} \right] \, ,
\label{eq822}
\eea
where $\al (m_Z) \approx \fr{1}{128}$%
~\cite{Tanabashi:2018oca}.

\begin{figure}[ht]
\centering
\begin{tabular}{cc}
\includegraphics[width=7.5cm]{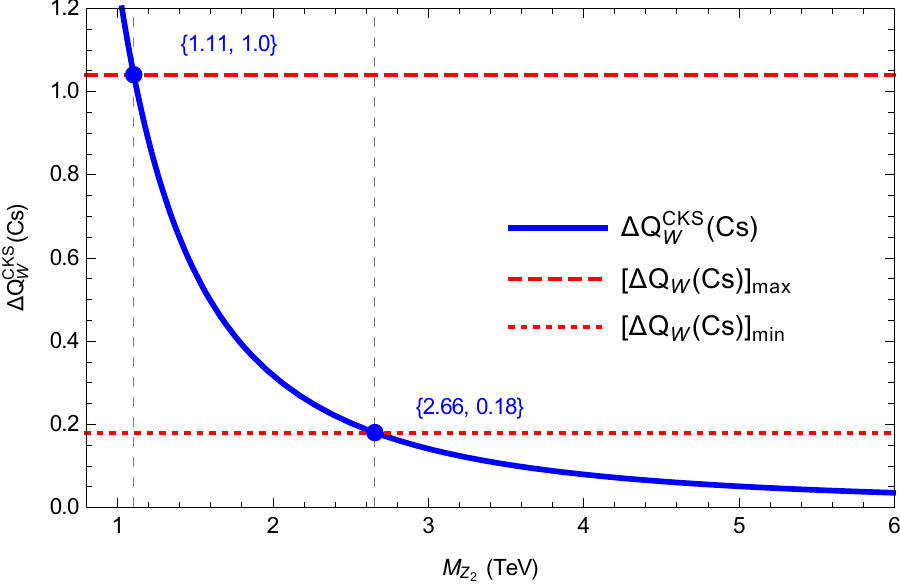}\hs&
\includegraphics[width=7.5cm]{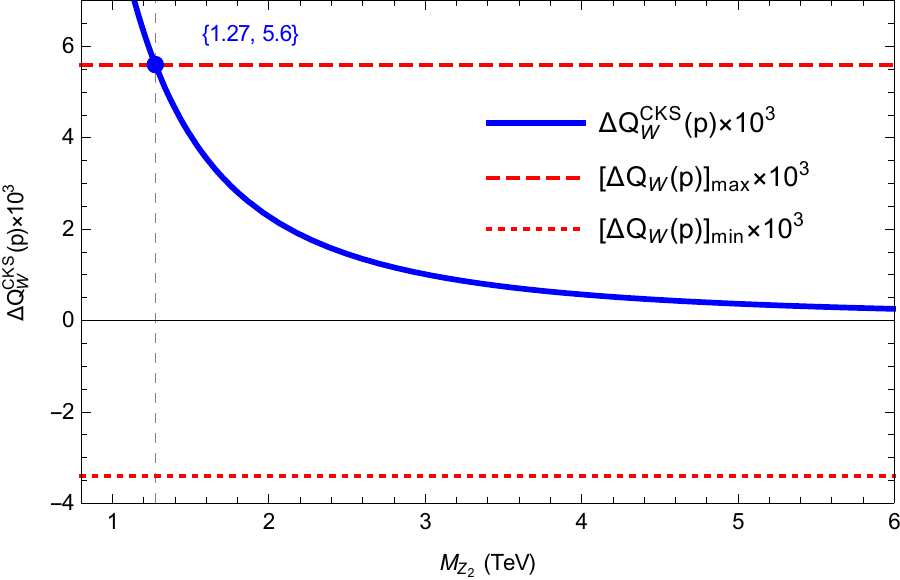} \\
\end{tabular}%
\caption{$\De Q_W^{\mathrm{CKS}}(Cs)$ and $\De Q_W^{\mathrm{CKS}}(p)$ as  functions of the $Z_2$ mass}
\label{fig1}
\end{figure}

In Fig. \ref{fig1}, we have plotted $\De Q_W^{\mathrm{CKS}}(Cs)$ and $\De Q_W^{\mathrm{CKS}}(p)$ as  functions of the extra neutral gauge boson $Z_2$ mass.
 It follows that the allowed values of the $Z_2$ mass is  $1.27 \, \mathrm{TeV}\le M_{Z_2}  \le 2.66$ TeV.
This range is less restrict than that from the $\rho$ data \cite{Long:2018dun} but it does not contradict it.

\section{Atomic parity violation in the
3 - 3 - 1 models for  arbitrary beta }
\label{331beta}

Let us briefly resume particle content of the model 3-3-1$\beta$~\cite{CarcamoHernandez:2005ka}.
 Here the $\beta$ is defined in Eq.~\eq{qoperator}.
The leptons lie in the $SU(3)_L$ triplet as follows
\be l_{a L} =\left( \nu_a \, , \hs e_a \, , \hs E^Q_a \right)^T \sim %
\left(1,3,-\fr 1 2 -\fr{\beta}{2\sqrt{3}} \right) \,  ,
\label{eq281}
\ee
 where $a=1, 2, 3$ is generation index.  This choice of lepton representation was called the model $F_2$ \cite{Buras:2014yna}. On the other hand,
 there exist models (model $F_1$) that $l_{a L}$ are antitriplets, but it can be shown that they are always  equivalent to some models
 with left-handed lepton triplets, in the sense that  both  have the same physics \cite{Martinez:2014lta, Hue:2018dqf}.
  Therefore, it is enough to focus on only the model $F_2$.

The chiral anomaly free requires the number of fermion triplets to be equal to that of fermion  antitriplets.  Therefore, in the model under consideration,
one generation of quarks transforms as $SU(3)_L$ triplets and two others transform as $SU(3)_L$ antitriplets. However, it is free to assign to quarks, provided
 the model is anomaly free.

   Here we adapt the notations in tables 1, 2, and 3 of Ref.~\cite{CarcamoHernandez:2005ka}.  In particular,  we consider the models containing  just
 three Higgs triplets defined in Refs.~\cite{CarcamoHernandez:2005ka,Buras:2012dp}, for example those given in Table 3
 of Ref. \cite{CarcamoHernandez:2005ka}.
  There are three different left-handed quark assignments, where the third, second or first  left-handed quark family
  is assigned as triplet, three respective models reps. A, B, and C were  introduced in Table 2
 of Ref. \cite{CarcamoHernandez:2005ka}.  Recall that the right-handed fermions are  $SU(3)_L$ singlets.

Note that the VEV of $\chi$ triplet provides masses of new particles, namely the exotic quarks and lepton as well as new gauge
bosons: $Z^\prime$ and bilepton gauge bosons $X$ and $Y$. Remember that the bottom element  of  $\chi$  does not carry lepton number,
 while the similar elements of $\eta$ and $\rho $ triplets  have lepton number equal to two. This means that only scalar components without
  lepton number can have VEV. In practice, to make the charged Higgs bosons having the integer value of electric charge, the parameter
  $\beta$ can take some special values only.

The masses  and mixing of the  neutral  gauge bosons  are presented in appendix \ref{APV331}. The needed gauge
couplings used to determine $Q_W$ are given in Table \ref{table_gV}, where only two models A and C
with different assignments
  of the first quark family are considered. The two models A and B have the same assignments of the first quark
   family, leading to the same APV result.
  Similar couplings were also given in Table 4 of Ref.~\cite{CarcamoHernandez:2005ka}, but they are different from
  ours by opposite signs, because of the difference choice of the phase of the $Z'$ state.
\begin{table}[htb]
	\caption{Vector and axial-vector coupling constants relevant for APV of  the $3-3-1\, \beta$
		model}
	\resizebox{16cm}{!}{
		\centering
		\begin{tabular}{|c|c|c|}
			\hline
			Standard Model & The  3 - 3 - 1 model (rep. A) &The 3 - 3 - 1 model (rep. C)\\
			\hline
$ g_A(e) =-\fr 1 2  $ & $g'_A(e) =   \fr{1 - (1 +\sqrt{3}\beta) s^2_W}{2\sqrt{3}\sqrt{1- (1+\beta^2)s^2_W}}$
&$g'_A(e) =  \fr{1 - (1 +\sqrt{3}\beta) s^2_W}{2\sqrt{3}\sqrt{1- (1+\beta^2)s^2_W}}$ \\
\hline
			$ g_V(u)=\fr 1 2 -\fr{4 s_W^2}3  $ & $g'_V(u) =
			\fr{-3+(3-5\sqrt{3}\beta )s_{W}^2 }{6\sqrt{3}\sqrt{1-(1+\beta
					^2 )s_{W}^2 }}$   & $%
			g'_V(u)  = \fr{ 3 - (3+5 \sqrt{3}\beta )s_{W}^2 }{6\sqrt{3}\sqrt{1-(1+\beta ^2 )s_{W}^2 }}$
			\\
			\hline
			$ g_V(d)=-\fr 1 2 +\fr{2s_W^2}3 $ &
			$g'_V(d)= \fr{-3 + (3 +\sqrt{3}\beta) s^2_W}{6\sqrt{3}\sqrt{1- (1+\beta^2)s^2_W}}  $& $g'_V(d)=
			\fr{ 3 - (3-\sqrt{3}\beta )s_{W}^2 }{6\sqrt{3}\sqrt{1-(1+\beta ^2 )s_{W}^2 }}$		
			\\
			\hline
	\end{tabular}}
	\label{table_gV}
\end{table}

Now we turn back to our main intention, namely the deviation of the weak charge $\De Q_W^{331}(Cs)$  in the $ 3-3-1\,\beta$ model.  The needed formula is
also  Eq.~\eq{del7}, will be applied to investigate the APV using the formula expressing the mixing $Z-Z'$ in terms of the model
 parameter $\beta$ \cite{Buras:2014yna}. The detailed steps to derive $\De Q_W^{331}(Cs)$  in the 3-3-1$\beta$ model are shown in appendix  \ref{APV331}.
Contribution from $\De \rho$ will be neglected. The relevant  $Z'$ couplings are given in Table~\ref{table_gV}.
 With  $M^2_{Z_1}\ll M^2_{Z_2}$, the $Z-Z'$ mixing angle $\phi$ can be formulated as follows \cite{Buras:2014yna}
	\be\label{eq-sphibe}
	s_\phi \simeq \tan \phi\simeq \fr{c_W^2}{3} \sqrt{f(\beta)}\left( 3\beta t^2_W +\sqrt{3} c_{2v}\right) \left[ \fr{M_Z^2}{M_{Z'}^2}\right],
	\ee
	where
		\begin{align}\label{eq_bbeta}
	f(\beta)=\fr{1}{1-(1+\beta^2)s_W^2}, \quad c_{2v}\equiv \cos(2\beta_v) =\frac{1-t^2_v}{1+t^2_v},
\quad t_v\equiv \tan\beta_v\equiv \fr{v_{\rho}}{v_{\eta}}.
	\end{align}
	In the numerical calculation, we will use $\fr{M_Z^2}{M_{Z'}^2}\simeq \fr{M_{Z_1}^2}{M_{Z_2}^2}$.
	
The parameter $t_v$ in Eq.~\eq{eq_bbeta} is  constrained from the Yukawa couplings of the top quark in the
third family, as in the well-known two Higgs doublet models (2HDM), for example see a review in Ref.~\cite{Branco:2011iw}. Depending on the model A (B, C), where
 left-handed top quarks are in triplets (anti-triplets),  they  get  tree level mass mainly from the coupling to $\eta$ ($\rho$) \cite{CarcamoHernandez:2005ka}.
  Especially, the top quark mass is $m_t\simeq \Ga^t \times \fr{v_{\rho(\eta)}}{\sqrt{2}}$, where the Yukawa coupling should satisfy the perturbative limit:  $|\Ga^t|<\sqrt{4\pi}$, resulting in a lower bound $v_{\rho(\eta)}>\fr{m_t}{\sqrt{4\pi}}$. As a consequence,  $t_v$ is constrained as
\be \label{eq_tvtriplet}
s_v=\fr{v_{\rho}}{\sqrt{v^2_{\rho} +v^2_{\eta}}}=\fr{gv_{\rho}}{2M_W} >\fr{g}{2M_W}\times \fr{m_t}{\sqrt{2\pi}}\simeq 0.28\Rightarrow t_v>t_0=\sqrt{\fr{1}{1-0.28^2}-1}\simeq 0.29
\ee
for top quark in anti-triplet (models B and C) and
\be \label{eq_tvantitriplet}
c_v=\fr{v_{\eta}}{\sqrt{v^2_{\rho} +v^2_{\eta}}}=\fr{gv_{\eta}}{2M_W} > 0.28\Rightarrow t_v<\sqrt{\fr{1}{0.28^2}-1}\simeq 3.43=t_0^{-1}
\ee
for top quark in triplet (model A). The constraint of $t_v$ in 3-3-1 models is  similar to the 2HDMs \cite{Branco:2011iw}.
We will use   $t_v\le 3.4$ for model A and $t_v\ge0.3$ for models B, C.

In the numerical investigation, we will look for  allowed regions satisfying three constraints of the APV data of Cs, the PVES data of proton, and the perturbative limit of Yukawa coupling of the top quark.  We will concentrate on the two models A and C.  The allowed regions predicted by model B will be addressed based on the weak charges predicted by the model A and the condition~ \eqref{eq_tvtriplet}. Numerical results are presented as follows.

\subsection{APV in the 3-3-1 model with $\bet = \pm \sqrt{3}$}
\subsubsection{The model with exotic leptons }
\begin{figure}[ht]
	\centering
	\begin{tabular}{cc}
		\includegraphics[width=7cm]{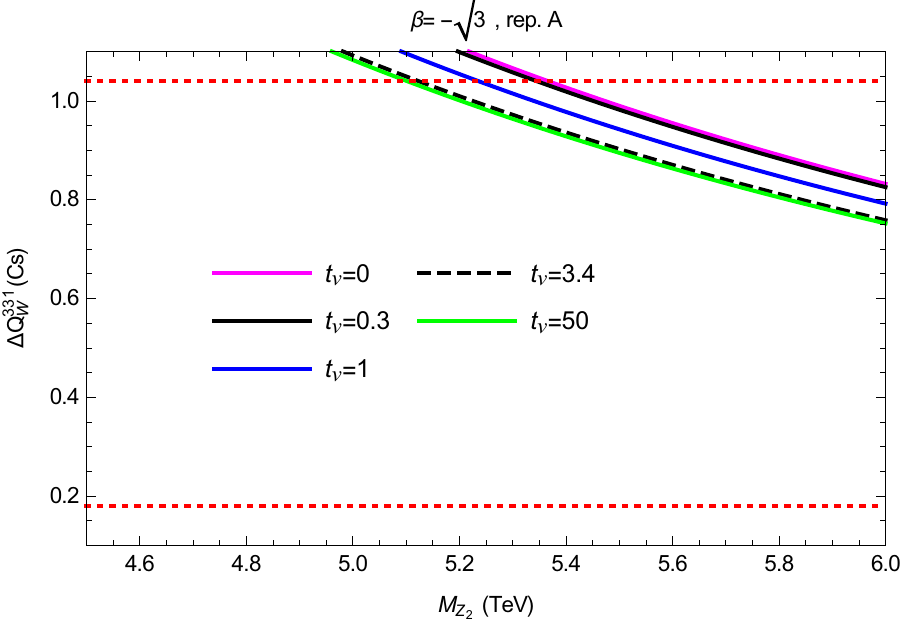} &
		\includegraphics[width=7cm]{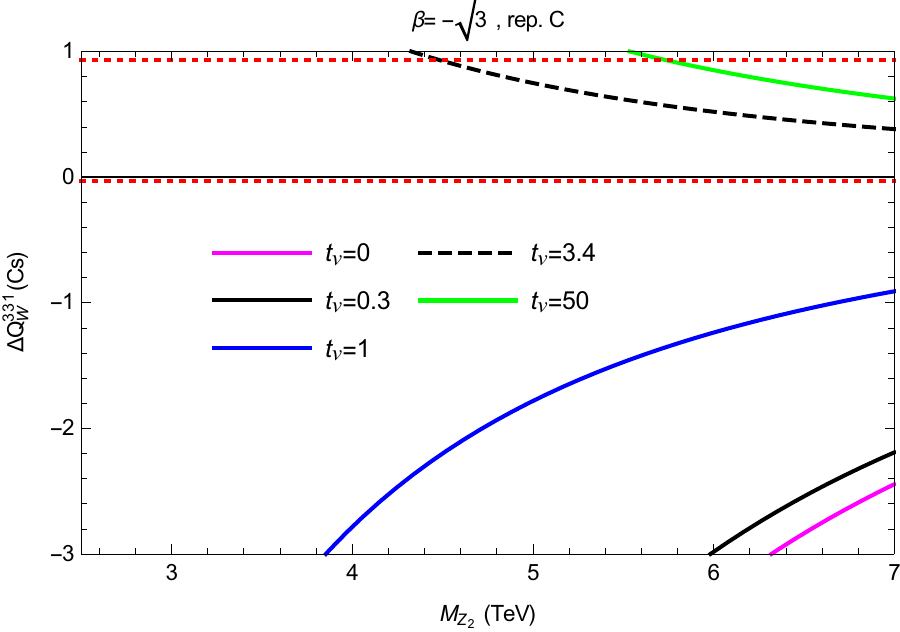} \\
		\includegraphics[width=7cm]{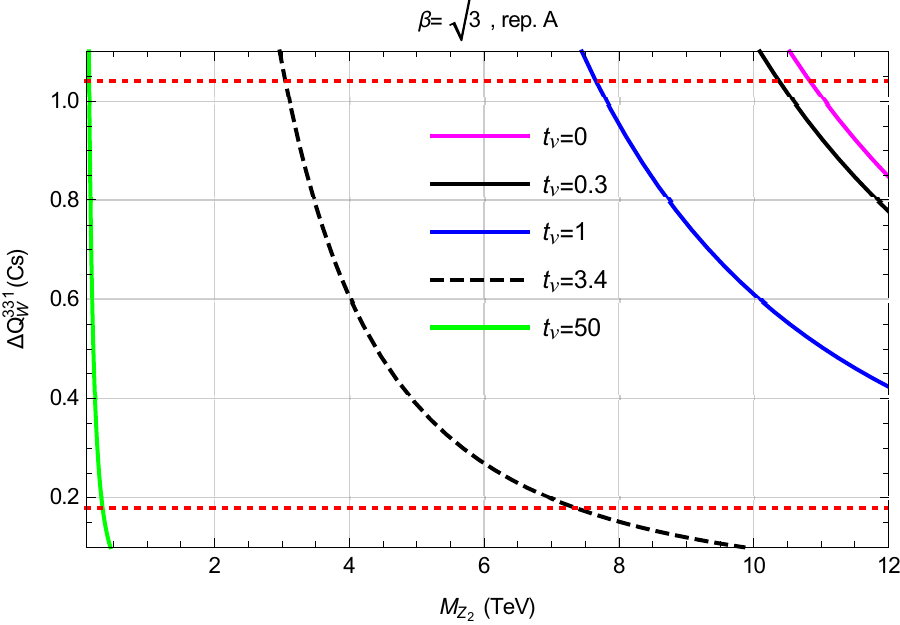} &
		\includegraphics[width=7cm]{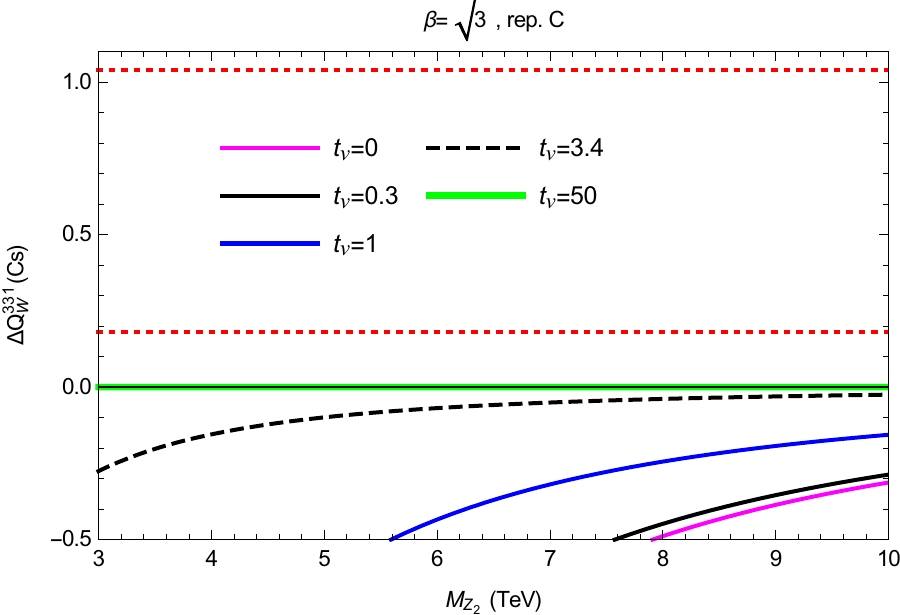}
	\end{tabular}%
	\caption{$\De Q_W^{331}(Cs)$ as a function of the $Z_2$ mass with $\beta= \pm \sqrt{3}$, predicted
		by  rep. A (C) in the left (right) panel. The two red dotted lines present two lower and upper experimental bounds of $\De Q_W(Cs)$. We use $s^2_{W}(M_{Z_2})=0.246$ and $g=0.636$ \cite{Buras:2013dea} }
	\label{fig_be3}
\end{figure}

The model we mention here is not the minimal 3-3-1 because the third components of
lepton triplets are the exotic ones. The numerical results are illustrated in  Fig.~\ref{fig_be3}.
We used the numerical values of the $SU(2)_L$ gauge boson couplings and the  Weinberg angle relating with $Z'$ given in
 Ref.~\cite{Buras:2013dea}, where the renormalization group evolutions  are taken into account. It also gives a consequence that the  limit for perturbative
   calculations requires $M_{Z_2}<4$ TeV.
 In the models under consideration, the relation  between $g_X$ and $g$ is determined by Eq.~\eq{eq_t}
from which the Landau pole arises at $s_W^2 = 1/(1+\beta^2)$. For the $\beta=\pm \sqrt{3}$, the models lose their perturbative
character at the scale around   4 TeV \cite{Ng:1992st, Frampton:2002st,Dias:2004dc, Martinez:2006gb,  Buras:2013dea}.
 We accept that the models will be ruled out if there are not any regions satisfying $ M_{Z_2}\le 4$ TeV.

From Fig. ~\ref{fig_be3}, we get the lowest value of $M_{Z_2}$ given in Table \ref{Bound3}.
\begin{table}[htb]
	\caption{
		Lower bounds of  of $M_{Z_2}$[TeV] with  $\beta= \pm \sqrt{3}$ from APV data of Cs}
\begin{tabular}{cc}
$\beta=-\sqrt{3}$	& $\beta=+ \sqrt{3}$\\
\begin{tabular}{c| c| c| c| c| c}
\hline
$t_v$&  $0$& $0.3 $ &$1$  & $3.4$ & $50$ \\
\hline
 A  & 5.37 &  5.35 & 5.24 & 5.12 & 5.10 \\
 C&  Excl.& Excl. & Excl. &4.24  & 5.43 \\
\hline
\end{tabular}
		\quad  &\quad
		 \begin{tabular}{c| c| c| c| c| c}
\hline
$t_v$&  $0$& $0.3 $ &$1$  & $3.4$ & $50$ \\
\hline
 A& 10.84& 10.38 & 7.66  &3.05 &0.14  \\
 C& 	Excl. &  Excl.&Excl.  &Excl.  &Excl. \\
\hline
\end{tabular}
\end{tabular}
\label{Bound3}
\end{table}	
%
The following remarks are in order:
\ben
\item For $\beta = - \sqrt{3}$,
 the model rep. A always predicts the lowest allowed  value of $M_{Z_2}$ around 5 TeV, where the perturbative property of the model is lost.
The same conclusion for the
model rep. C for $t_v = 50$ or $t_v\le1$.

\item For $\beta = + \sqrt{3}$,  the model rep.  C is excluded for all values of $t_v$.

\begin{figure}[ht]
	\centering
	\begin{tabular}{cc}
		\includegraphics[width=8cm]{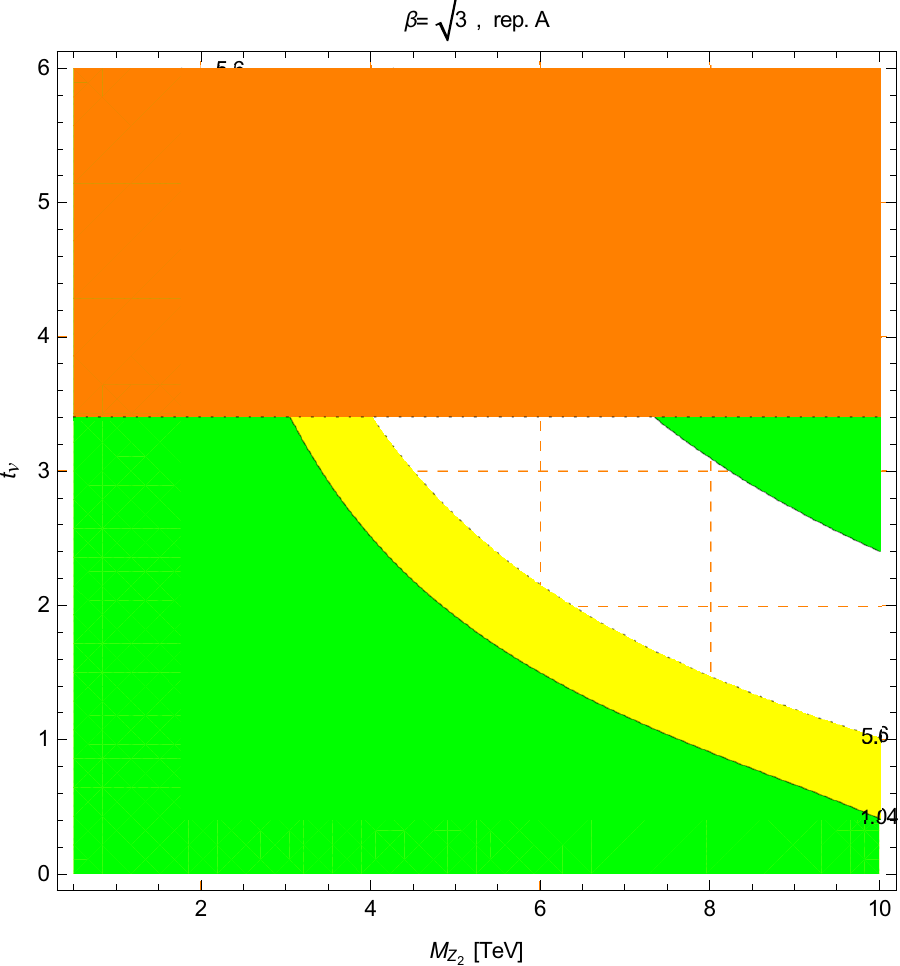}
		&
		\includegraphics[width=8cm]{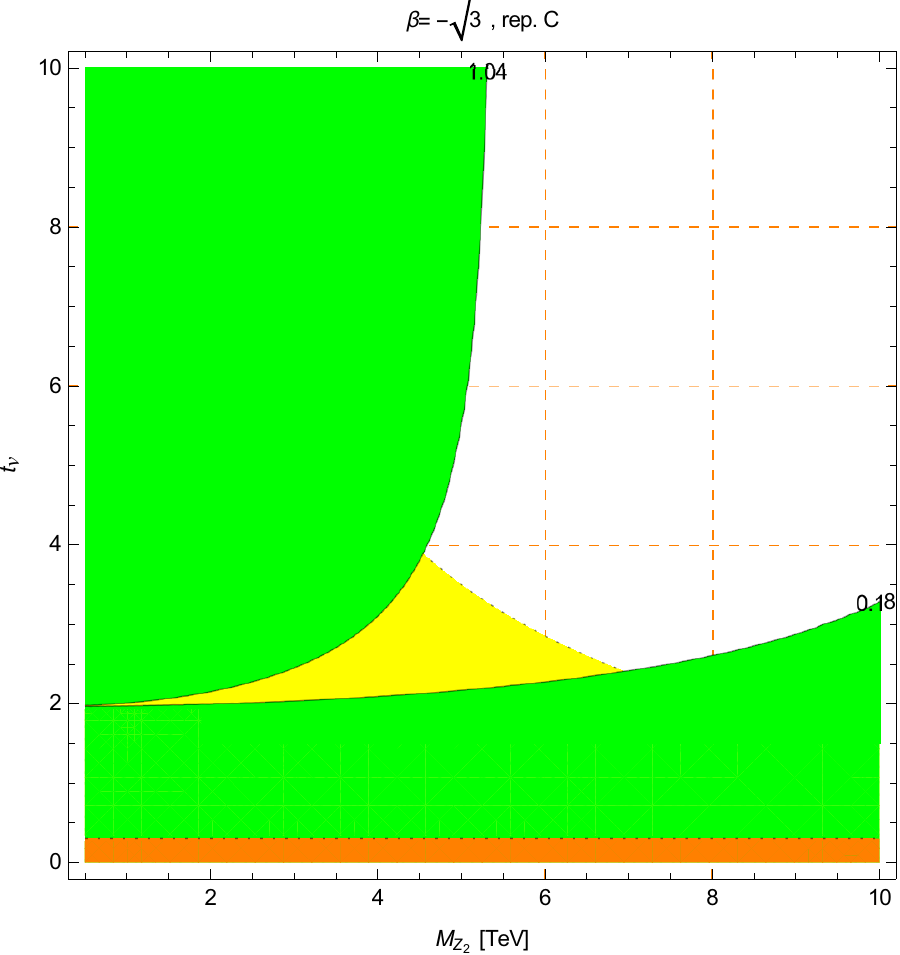}
	\end{tabular}%
	\caption{Allowed regions in the plane $M_{Z_2}-t_v$ predicted by  rep. A (C) with $\beta=\sqrt{3}$ ($\beta=-\sqrt{3}$), where the orange region is excluded by $t_v\le3.4$ ($t_v\ge0.3$). The green and yellow regions are  excluded by the APV data of Cs and PVES data of the proton, respectively.}
	\label{fig_contoursACpm3}
\end{figure}

\item  The value $t_v = 3.4$ is survived  for two models:  rep. C with $\beta= -\sqrt{3}$ ($M_{Z_2} \ge 4.24$)  and  rep. A with $\beta= \sqrt{3}$ ($M_{Z_2} \ge 3.05$).  Combining with the condition of the Yukawa coupling of top quark and PVES  data of proton, the allowed regions are more strict, see Fig.~\ref{fig_contoursACpm3}.
The $M_{Z_2}$ values must satisfy $M_{Z_2}\ge 4$ TeV for model rep. A and $M_{Z_2}\ge 4.5$ TeV for model rep. C. Hence the lower bounds from combined data are more constrained than those obtained from the data of APV of Cs alone.
\een

\subsubsection{The minimal 3-3-1 model}
Apart from the  case of $\beta= - \sqrt{3}$ mentioned above,  another model with $\beta= - \sqrt{3}$ but no new charged lepton, i.e. the third components of the lepton triplets are conjugations of right-handed SM charged leptons, is well known as  the minimal 3-3-1 model (M331).  The gauge couplings relevant to  the APV are given in table~\ref{table_gVM331}.

\begin{table}[htb]
	\caption{Vector and axial-vector coupling constants relevant for APV M331 models}
	\resizebox{13cm}{!}{
		\centering
		\begin{tabular}{|c|c|c|}
			\hline
			Standard Model & rep. A &  rep. C\\
			\hline
			$ g_A(e) =-\fr 1 2  $ & $g'_A(e) =  - \fr{\sqrt{1-4 s_W^2}}{2\sqrt{3}}$
			&$g'_A(e) =- \fr{\sqrt{1-4 s_W^2}}{2\sqrt{3}} $ \\
			\hline
			$ g_V(u)=\fr 1 2 -\fr{4 s_W^2}3  $ & $g'_V(u) = \fr{-1 +6s_W^2}{2\sqrt{3}\sqrt{1-4 s_W^2}}$
			& $ g'_V(u)  =   \fr{1 +4 s_W^2}{2\sqrt{3}\sqrt{1-4 s_W^2}}$
			\\
			\hline
			$ g_V(d)=-\fr 1 2 +\fr{2s_W^2}3 $ &
			$g'_V(d)=  -\fr{1}{2\sqrt{3}\sqrt{1-4 s_W^2}}$& $g'_V(d)=  \fr{1 -2s_W^2}{2\sqrt{3}\sqrt{1-4 s_W^2}}$		
			\\
			\hline
	\end{tabular}}
	\label{table_gVM331}
\end{table}

\begin{figure}[th]
	\centering
	\begin{tabular}{cc}
		\includegraphics[width=7cm]{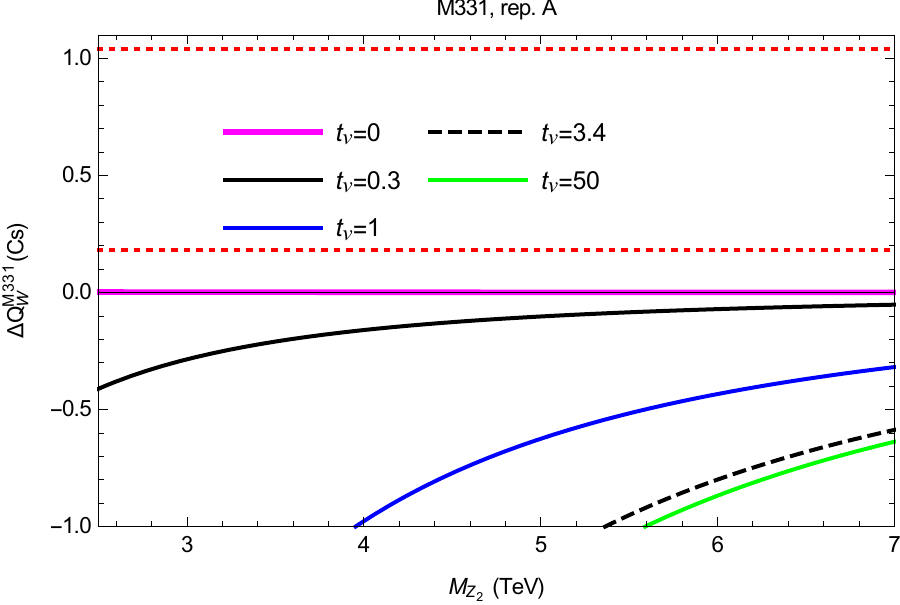} &
		\includegraphics[width=7cm]{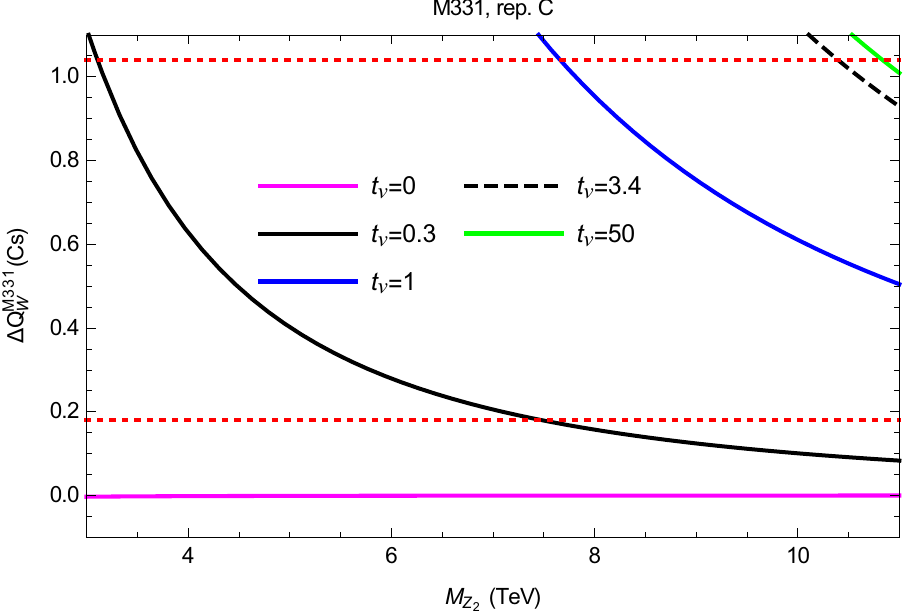}
	\end{tabular}%
	\caption{$\De Q_W^{\mathrm{M331}}(Cs)$ as a function of the $Z_2$
		mass, predicted by the M331 model with the case  of   rep. A (C) in the left (right) panel. We have used $s^2_{W}(M_{Z_2})=0.246$ and $g=0.636$ \cite{Buras:2013dea}}
	\label{fig_M331}
\end{figure}

The numerical results are illustrated in  Fig.~\ref{fig_M331}.
 It can be seen that all curves are out of the allowed range given by experiment in the framework of rep. A. In contrast,
 there still exist allowed $M_{Z_2}$
  values in the rep. C. Furthermore, small allowed $M_{Z_2}$  corresponds  to small $t_v$. Some specific limits are summarized in Table~\ref{BoundM331}.
\begin{table}[htb]
	\caption{
	Allowed range of  $M_{Z_2}$  for  M331}
	\begin{tabular}{c| c| c| c |c |c }
		\hline
		$t_v$&  $0$& $0.3 $ &$1$  & $3.4$ & $50$ \\
		\hline
		A& excl. & excl.   & excl.  &excl.  &excl.\\
		C& excl.& [3.11, 7.47]&[7.66, 18.41]& [10.40, 24.99] & [10.83, 26.04] \\
		\hline
	\end{tabular}	\label{BoundM331}
\end{table}

\begin{figure}[ht]
	\centering
	\begin{tabular}{cc}
		&
		\includegraphics[width=8cm]{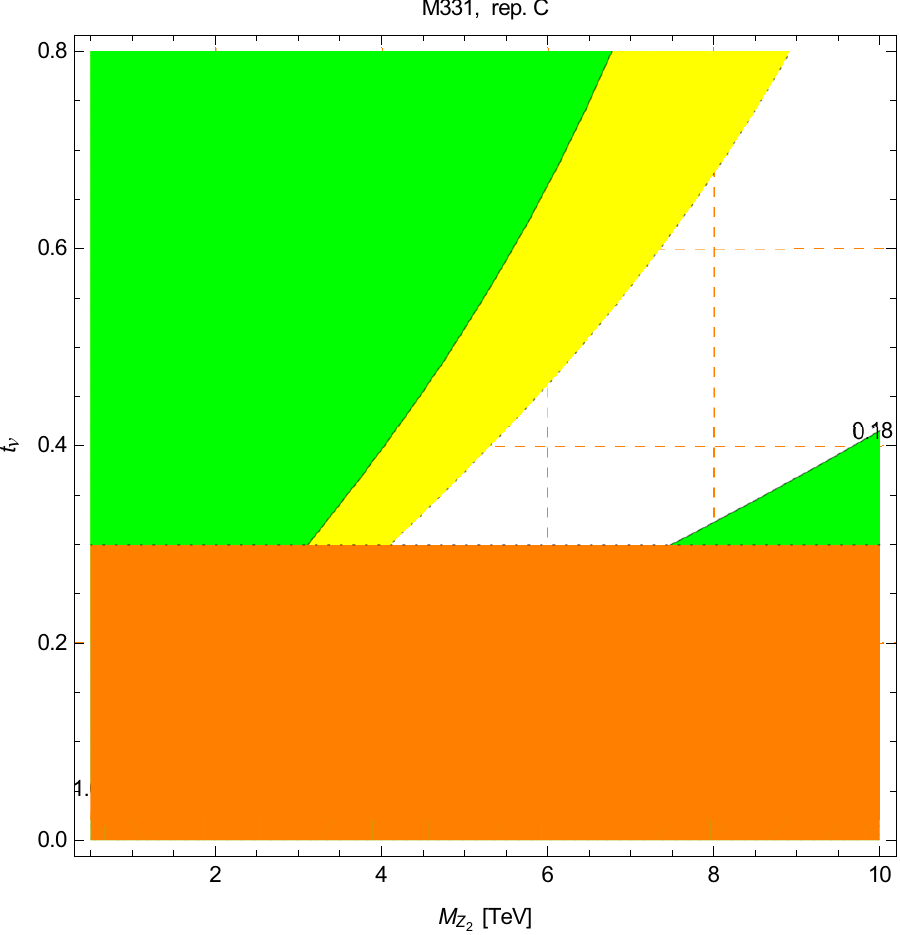}
	\end{tabular}%
	\caption{Allowed regions in the plane $M_{Z_2}-t_v$  predicted by  rep. C  of the M331. The orange region is excluded by $t_v\ge0.3$. The green and yellow regions are  excluded by the APV data of Cs and PVES data of the proton, respectively.}
	\label{fig_contoursMCpm3}
\end{figure}

We see that the data on APV of Cesium excludes the M331 model with rep. A, but still allows rep. C with some small $t_v$, for example $m_{Z_2}\ge 3.11$ TeV with $t_v=0.3$. Combining with the conditions of $t_v\ge0.3$ and the PVES data of proton will give a more strict lower bound $m_{Z_2}\ge 4$ TeV, see  Fig.~\ref{fig_contoursMCpm3}. The lower bound of $M_{Z_2}$ obtained from the PVES data of proton is more strict than the APV data of Cs.

\subsection{APV in the 3-3-1 model with $\bet = \pm  \fr{1}{\sqrt{3}}$}
\begin{figure}[ht]
	\centering
	\begin{tabular}{cc}
		\includegraphics[width=7cm]{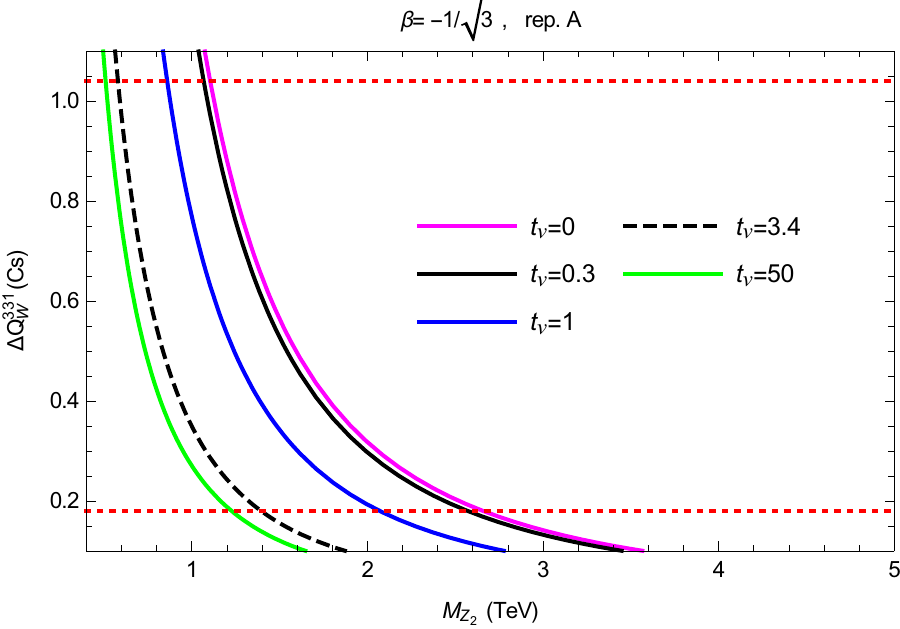} &
		\includegraphics[width=7cm]{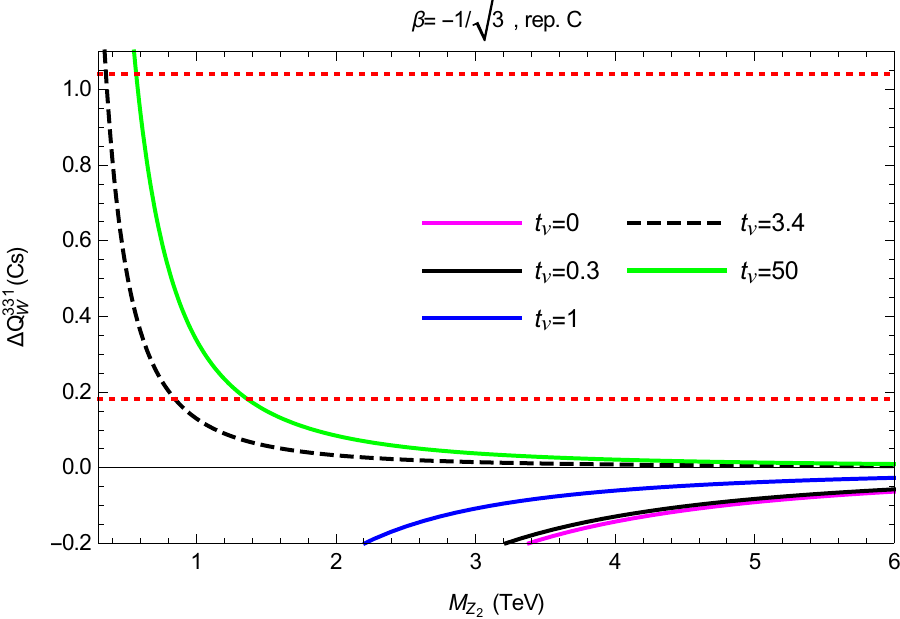} \\
		\includegraphics[width=7cm]{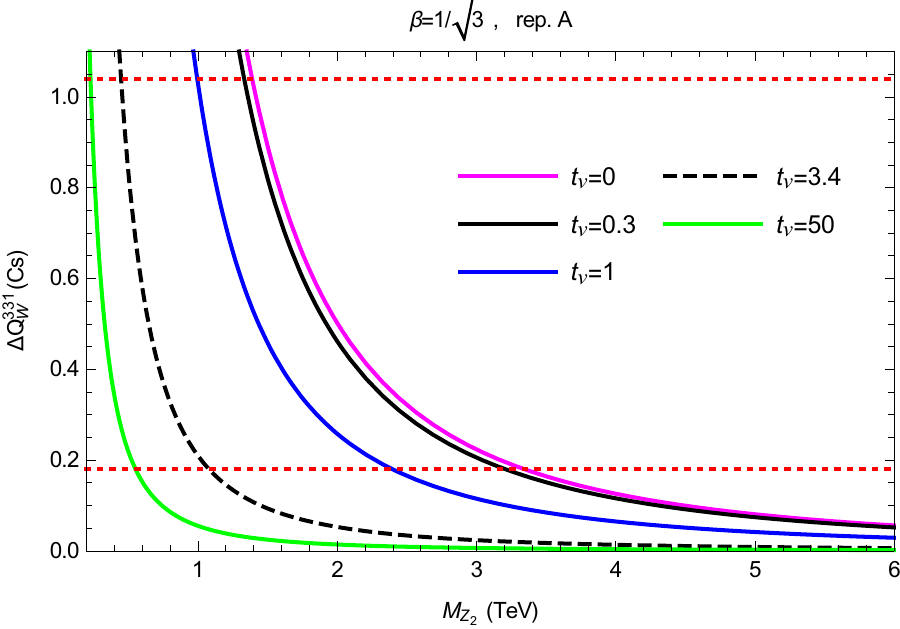} &
		\includegraphics[width=7cm]{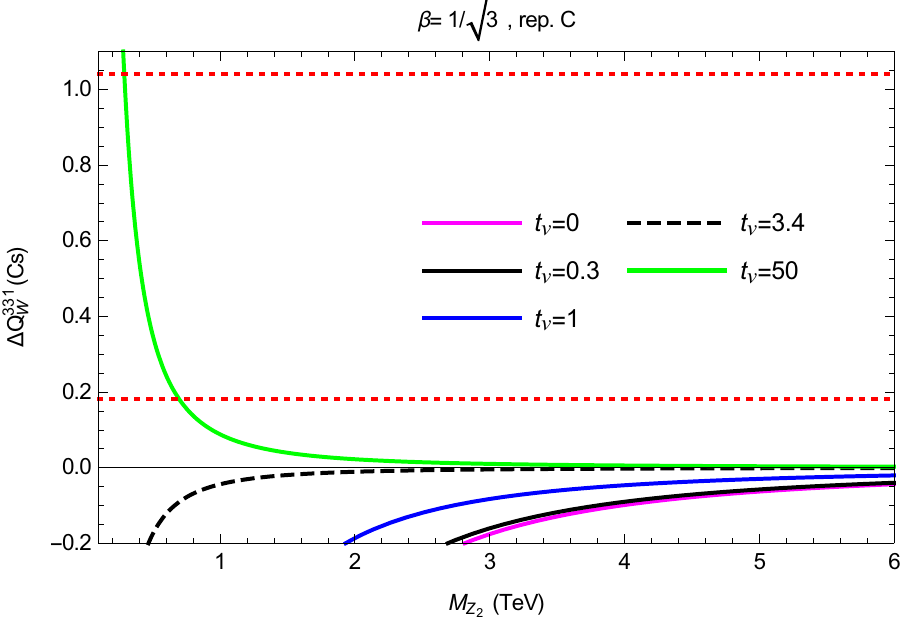}
			\end{tabular}%
	\caption{$\De Q_W^{331}(Cs)$ as a function of the $Z_2$ mass with $\beta= \pm \fr{1}{\sqrt{3}}$, predicted by
	rep. A (C) in the left (right) panel.}
	\label{fig_be1}
\end{figure}

  Regarding the couplings of $Z'$ at  $M_{Z'}=\mathcal{O}(1)$ TeV, we will  use
 $g(M_{Z_2})=0.633$, $s^2_W(M_{Z_2})=0.249$ for $\beta=0,\pm\fr{1}{\sqrt{3}},\pm\fr{2}{\sqrt{3}}$
 \cite{Buras:2013dea, Buras:2016dxz, Buras:2014yna}.  The numerical results obtained from APV of Cesium  are shown in Fig.~\ref{fig_be1}.
 Some limits for $\beta= \pm \fr{1}{\sqrt{3}}$ are explicitly  presented in Table~\ref{Bound1can3}.
\begin{table}[htb]
	\centering
	\caption{ Allowed range of  $M_{Z_2}$[TeV]  for  $\beta= \pm\fr{1}{ \sqrt{3}}$ }
	\begin{tabular}{c|c}
\hline
		$\beta=- \fr{1}{ \sqrt{3}}$	&	\begin{tabular}{c | c| c| c| c| c}
			$t_v$&  $0$& $0.3 $ &$1$  & $3.4$ & $50$ \\
			\hline
			A  & [1.11, 2.66] & [1.06, 2.57]& [0.86, 2.07] & [0.58, 1.39]& [0.51, 1.23] \\
			C& Excl. & Excl. & Excl. & [0.35, 0.85] & [0.57, 1.37]\\
		\end{tabular}\\
\hline
	 $\beta=+ \fr{1}{ \sqrt{3}}$
	 &
		\begin{tabular}{c| c| c| c| c| c}
			A& [1.39, 3.34] &[1.33, 3.20]  &[1.00, 2.39]  &[0.45, 1.08]  & [0.23, 0.55] \\
			C& Excl. & Excl. & Excl.& Excl.& [0.29, 0.70]\\
		\end{tabular}\\
\hline
	\end{tabular}
	\label{Bound1can3}
\end{table}	
One gets the following results
\ben
\item For both $\beta=\pm \fr{1}{\sqrt{3}}$, the model rep. A survives with all $t_v$. The allowed values of $M_{Z_2}$  decrease with increasing $t_v$.
\item The model C survives with only large  $t_v$ and small $M_{Z_2}\le 1.5$ TeV.
\een

\subsection{APV in the 3-3-1 model with $\bet = 0$}

\begin{figure}[ht]
	\centering
	\begin{tabular}{cc}
		\includegraphics[width=7cm]{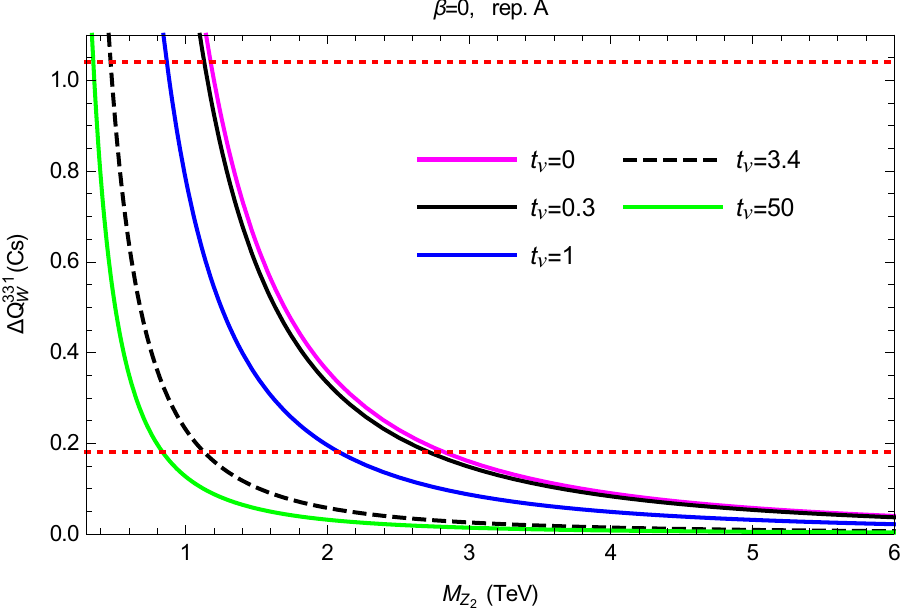} &
		\includegraphics[width=7cm]{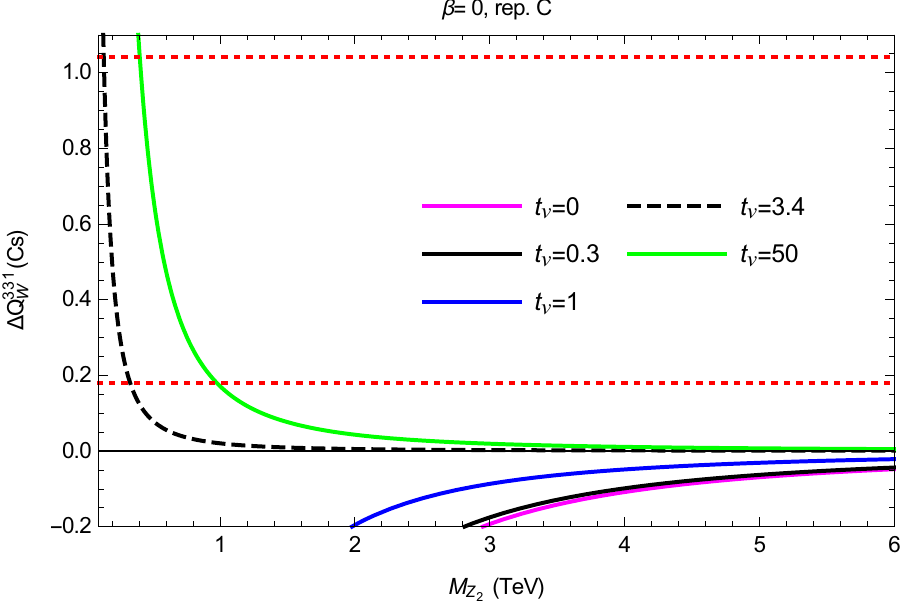}
	\end{tabular}%
	\caption{$\De Q_W^{331}(Cs)$ as a function of the $Z_2$ mass with $\beta= 0$, predicted by  rep. A (C) in the left (right) panel.}
	\label{fig_be0}
\end{figure}

The 3-3-1 model with $\beta = 0$ has been recently constructed in Ref.~\cite{Hue:2015mna}. The numerical results are shown in the Fig.~\ref{fig_be0}.
Result is summarized in Table~\ref{Boundm0}.
\begin{table}[htb]
		\caption{
Allowed range of $M_{Z_2}$ for  $\beta=0$}
		\begin{tabular}{c|c |c | c| c| c}
		\hline
		$t_v$&  $0$& $0.3$ &$1$  & $3.4$ & $50$ \\
		\hline
		A& [1.18, 2.83] & [1.13, 2.72]   & [0.87, 2.09]  & [0.47, 1.13]  & [0.35, 0.84]\\
		C& 	excluded & excluded & excluded & [0.14, 0.33]& [0.41, 0.98]\\
		\hline
	\end{tabular} 	\label{Boundm0}
\end{table}
We see the similarity to  the cases $\beta= \pm \fr{1}{ \sqrt{3}}$. These models predict  a rather light $M_{Z_2}$, which was  mentioned previously in other models \cite{Komachenko:1989qn,Boucenna:2016qad, Hue:2016nya, He:2017bft}.  The difference is that the allowed ranges of $M_{Z_2}$ drift increasingly for  $\beta$ changing from $-\frac{1}{\sqrt{3}}$ to $\frac{1}{\sqrt{3}}$.

\begin{figure}[ht]
	\centering
	\begin{tabular}{cc}
		&
		\includegraphics[width=7cm]{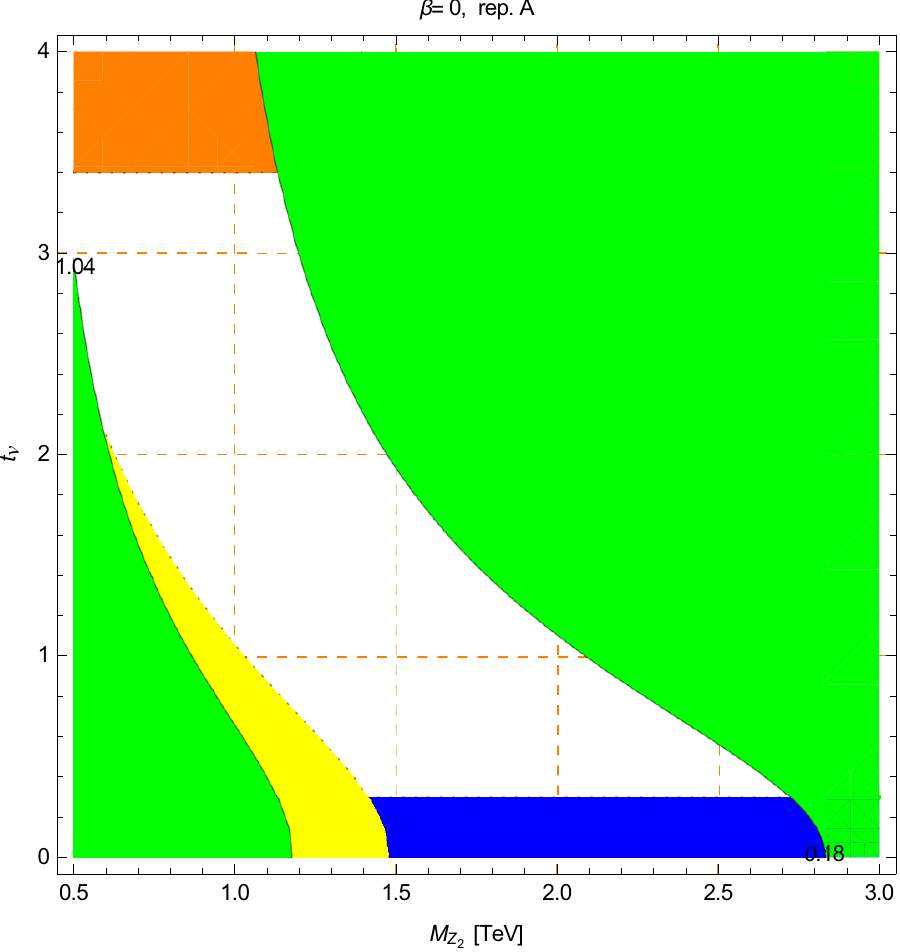}
	\end{tabular}%
	\caption{Allowed regions in the plane $M_{Z_2}-t_v$, predicted by  rep. A with $\beta=0$.
	The orange, green and yellow regions are  excluded by the condition $t_v\le 3.4$, the APV data of Cs and the PVES data of the  proton, respectively. The blue region is excluded by the condition $t_v\ge 0.3$.}
	\label{fig_contoursMA}
\end{figure}

There are some common properties for model rep. A,  that we can see from all the above plots.  Namely,  the lower bounds of $M_{Z_2}$
 involved with the APV of Cs are always increased corresponding to the decreasing $t_{v}$. As a result,
  an illustration of the allowed regions  is shown in Fig.~\ref{fig_contoursMA} for $\beta=0$.
 Hence, perturbative condition $t_v\le 3.4$ excludes regions of small $M_{Z_2}$. In the region with
 small $t_v\rightarrow 0$, the PVES data of the proton gives more strict lower bounds than the APV data of Cesium,
 see again Fig.~\ref{fig_contoursMA}.  The largest allowed values of $M_{Z_2}$ is around 2.8 TeV.
 It increases to 4.65 TeV for $\beta=\fr{2}{\sqrt{3}}$.

 Regarding to model rep. B, which has the same results of APV, but the allowed regions satisfy $t_v>0.3$,
 which can be seen in  Fig.~\ref{fig_contoursMA}. The model B excludes regions containing large $M_{Z_2}$.

\begin{figure}[ht]
	\centering
	\begin{tabular}{cc}
	\includegraphics[width=7cm]{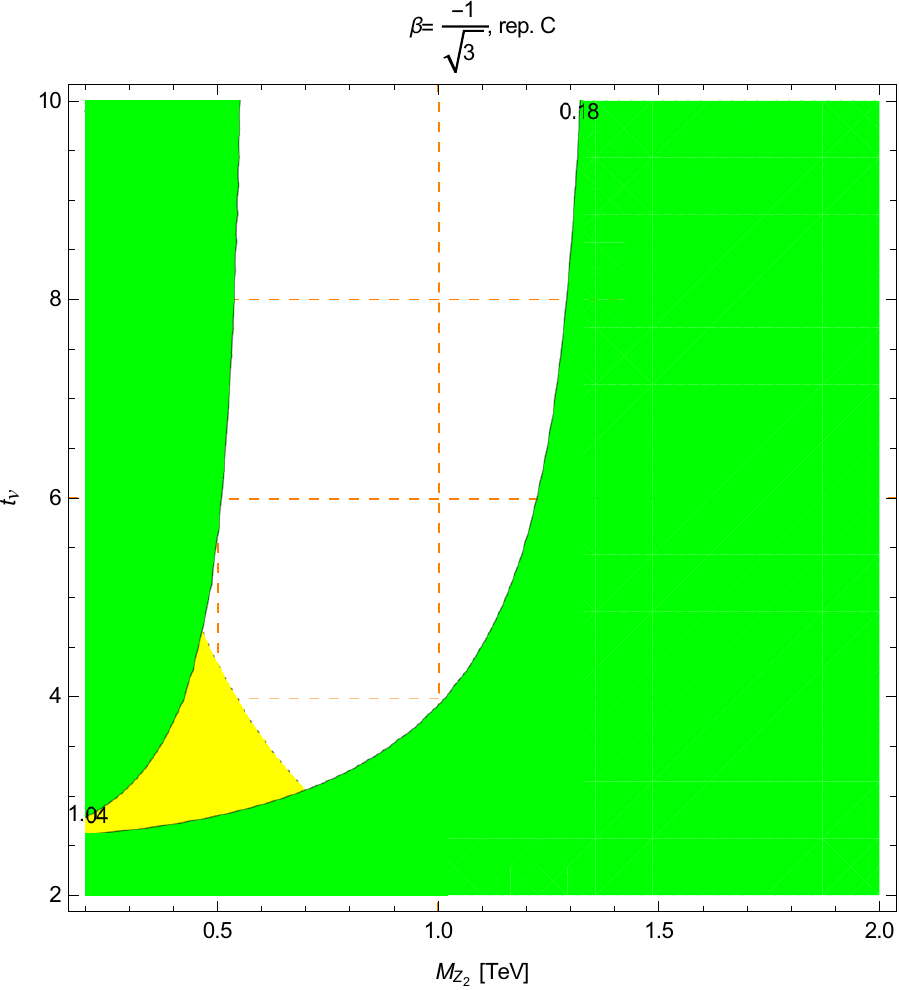} &
		\includegraphics[width=7cm]{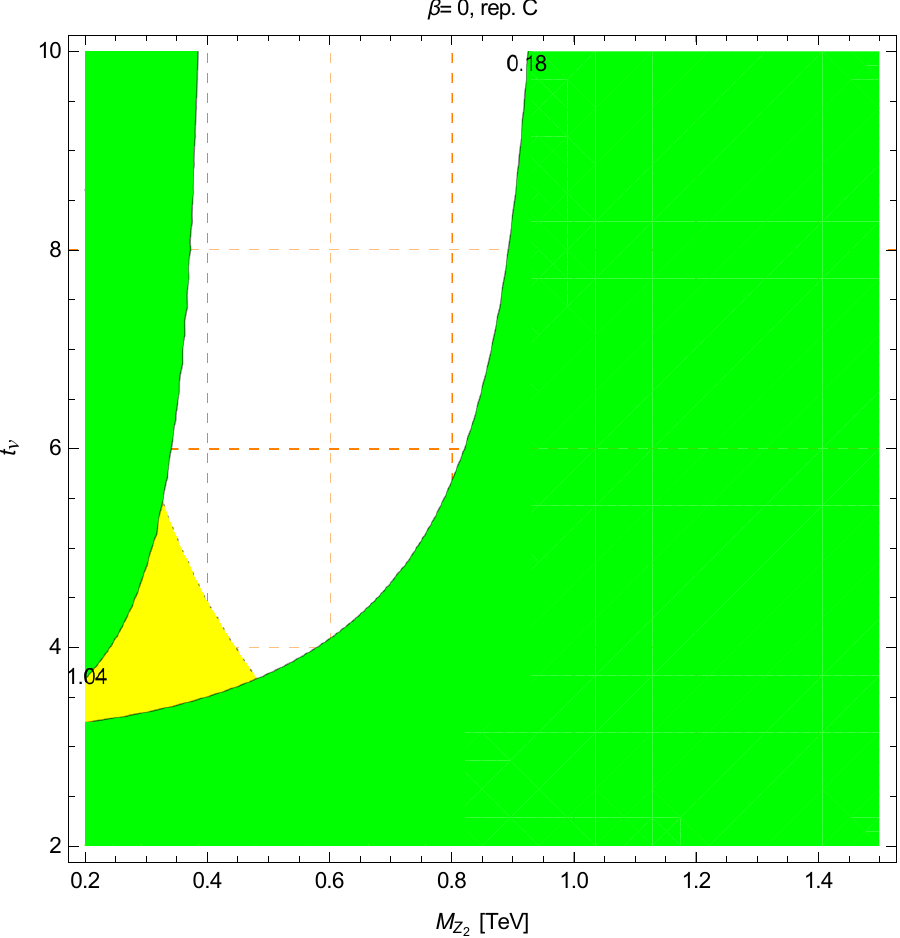}
	\end{tabular}%
	\caption{Allowed regions in the plane $M_{Z_2}-t_v$, predicted by  rep.C.  The green and yellow regions are
	 excluded by the APV data of Cs and PVES data of the  proton, respectively.}
	\label{fig_contoursMCpm1}
\end{figure}

 Illustrations for allowed regions predicted by model rep. C with $\beta=0,-\fr{1}{\sqrt{3}}$ are shown in Figs.~\ref{fig_contoursMCpm1}.
The PVES data of the proton excludes small $m_{Z_2}$ and $t_v$, which is more strict that the pertubative limit
 of top quark coupling. The allowed  regions predict only small values $M_{Z_2}<1.5$ TeV. For other $\beta$
 satisfying $|\beta|<\sqrt{3}$, the situations are similar to the mentioned illustrations, but
  the  upper bounds of $M_{Z_2}$ may reach larger value of $2.5$ TeV.

\section{Conclusions}
\label{conclusion}
The effects of the weak charges of  Cesium and the  proton on the parameter spaces of 3-3-1 models are discussed under the current experimental APV and PVES data and
 the perturbative limit of the Yukawa coupling of the top quark.
 Within a recently  proposed 3-3-1 CKS, we get the lowest value of $M_{Z_2}$ to be 1.27 TeV. This limit is slightly lower than that concerned from the LHC searches, $B$ decays or $\rho$ parameter data.

  We have also performed studies for the other versions of the 3-3-1 models with three Higgs triplets. Here are the main conclusions:
\begin{itemize}
	\item $\beta=\pm \sqrt{3}$, the regions with $M_{Z_2}<4$ TeV are excluded in the frameworks of all models reps.  A, C and M331. They are ruled out when the perturbative calculation limit are required, where the Landau pole of the models  happens at the scale
	around 4 TeV.  The APV data of Cesium alone rules out only three cases of the model C with $\beta=-\sqrt{3}$,  the model rep. A with $\beta=\sqrt{3}$, and the M331 rep. A. Other cases are ruled out based on the PVES data of proton and top quark couplings limit.
	
	\item For $|\beta|<\sqrt{3}$, for example $\beta=0,\pm\fr{1}{\sqrt{3}}$, the   allowed regions are affected significantly by the PVES data of proton, namely it results in the lower bounds of $M_{Z_2}$ more strict than those obtain from the APV data of Cesium.  This point was not mentioned previously.
	\item For $\beta=0,\pm\fr{1}{\sqrt{3}}$, the model rep. C favors the regions with only small $M_{Z_2}<1.5$ TeV.
	
\item For $\beta=\pm\fr{1}{\sqrt{3}}$, the model reps. A gives larger allowed values of $M_{Z_2}$. This model will not be ruled out by other constraints from LHC, where $M_{Z_2}\ge2.5$ TeV with assumption that $Z_2$ does not decay to heavy fermions \cite{Richard:2013xfa, Salazar:2015gxa}, or all heavy fermion masses are 1 TeV \cite{Coutinho:2013lta}.   A reasonable lower bound were acceptable in literature $M_{Z_2}\ge 1$ TeV \cite{Buras:2013dea, Hue:2017lak}.

\item The model rep. B also survives,  although the perturbative limit of the Yukawa couplings of the top quark gives  constraints on the allowed regions with large $M_{Z_2}$.
\end{itemize}

From our discussion, we emphasize that the information of PVES data of proton and the pertubative limit of top quark Yukawa couplings  are  as  important as that obtained from the APV of Cesium, therefore all of them should be discussed simultaneously to constrain the parameter space of the 3-3-1 models.  The numerical calculations have also shown that the allowed regions predicted by the two models reps. B and C disfavor the large $M_{Z_2}$ hence they may be ruled out by future constraints  from colliders such as LHC, especially the model rep. C. While the model rep. A may still be survived, resulting in that the heaviest  quark family must treat differently from the remaining. Furthermore,  our work concerns that the  improved weak charge data from the future experiments  will be important  to decide which quark family in realistic 3-3-1 models should be assigned differently from the two remaining families.

The  recent data of APV and PVES  is consistent with the data on the mass  difference of neutral meson \cite{Long:1999ij} in the sense that the third
family should be treated differently  from the first twos. This also  gives a reason why the top quark is so heavy.

\section*{Acknowledgments}

LTH thanks Le Duc Ninh for interesting discussions and recommending Ref.~\cite{Martinez:2014lta}.  We thank prof. Maxim Klopov for communicating with us. Especially, we are grateful Prof.   David Armstrong for explaining the PVES experiment measuring the proton's  weak charge. We acknowledge the financial support of the International Centre of Physics at the Institute of Physics, Vietnam Academy of Science and Technology. This research has been financially supported by the Vietnam National Foundation for
Science and Technology Development (NAFOSTED) under grant number
103.01-2017.356.
\appendix
\section{ Derivation of the weak charge expression in the models with extra neutral gauge boson}
\label{deviation}
Nowadays, a lot of  beyond Standard Models  contain extra neutral gauge bosons associated with new diagonal generators such as $T_8, T_{15}$
or extra generators of the new $U(1)_N$ groups. The above mentioned neutral gauge bosons  will give contribution to the
 atomic parity violation. So we will provide a detailed analysis of the APV in the light of extra gauge bosons.

Some authors use the notations with different coefficients and signs associated with axial part ($\ga_5$). Here we point out
the relation among the notations.
\subsection{Notations}

For convenience to apply the results into our calculations, we review here more detailed steps to derive analytic formulas of $\De Q_W(^A_ZX)$.
However, we firstly consider the case with just one extra neutral gauge boson $Z^\prime$. After some steps of diagonalization
in neutral gauge boson sector, we come to two states $Z$ and $Z^\prime$ with
  Lagrangian \eq{eqVff}.
It is emphasized that the $Z$ and $Z^\prime$ are mixed and  the  physical states are a result of the last
step of diagonalization which is discussed latter. In conventional way, the $Z$ and $Z^\prime$ are mixing with an angle $\phi$; and the consequence is a pair
of the physical bosons $Z_1$ and $Z_2$.
Relations between the notations in Eq.~\eq{eqVff} and those mentioned in Ref.~\cite{Altarelli:1991ci} are
\be
g_V(f) = 2  v_f \, , \hs g_A(f)  = - 2  a(f) \, , \hs g'_V(f)  = 2  v'_f \, ,
 \hs g'_A(f) =  - 2  a'(f) \, .
\label{eq121t}
\ee

We will base on the approach to derive the deviation comparing with the results given by
 G. Altarelli  et al. \cite{Altarelli:1991ci}.
The equivalence of the neutral gauge boson states between our notation and those in
 Ref.~\cite{Altarelli:1991ci} are  \eq{eq121t} and
 \begin{equation*}
 (Z,Z')\equiv(Z_0,\,Z'_0),\quad (Z_1,Z_2)\equiv (Z,Z'),\quad \xi_0\equiv \phi, \quad \tilde{g}=g'=gt_W.
 \end{equation*}
The mixing matrix $O$ relating
 two base of neutral gauge bosons are:
\begin{align}\label{eqmixingZZ'}
O=\begin{pmatrix}
c_{\xi_0}& -s_{\xi_0} \\
s_{\xi_0}& c_{\xi_0}
\end{pmatrix} \equiv \begin{pmatrix}
c_\phi & -s_\phi  \\
s_\phi & c_\phi
\end{pmatrix},
\end{align}
which give $(Z_1,Z_2)^T=O(Z,Z')^T$.
 We will use our notations in the following calculations.

Lagrangian containing gauge couplings of neutral gauge bosons in the basis $(Z,\,Z')$ is
\begin{align}\label{eqZZpffb}
\mathcal{L}^{\mathrm{BSM}}_{Vff}&=J_\mu Z^\mu +J'_\mu Z'_\mu\crn
&\equiv  \fr{g}{2c_W}\sum_f \overline{f}\ga^\mu [g_V(f) -\ga_5 g_A(f)]f Z_\mu
+\fr{g}{2c_W}\sum_f \overline{f}\ga^\mu[g'_V(f) -\ga_5 g'_A(f)]f Z'_\mu\, .
\end{align}
 On the other hand,  in terms of physical
 neutral gauge boson mediations $Z_1$ and $Z_2$, this  Lagrangian can be written as follows
\begin{align}\label{eqZ12ffb}
\mathcal{L}^{\mathrm{BSM}}_{Vff}&= \fr{g}{2c_W}\sum_f \overline{f}\ga^\mu[g^{(1)}_V(f)
 - \ga^5 g^{(1)}_A(f)]f  Z_{1\mu} +\fr{g}{2c_W}\sum_f \overline{f}\ga^\mu[g^{(2)}_V(f)
 - \ga^5 g^{(2)}_A(f)]f  Z_{2\mu},
\end{align}
where the couplings $g^{(1)}_V(f)$, $g^{(2)}_V(f)$, $g^{(1)}_A(f)$ and $g^{(2)}_A(f)$ are gauge couplings of the physical states of neutral gauge boson, which
will be determined as functions of $g_{V,A}(f)$ and $g'_{V,A}(f)$.
Eq.~\eq{eqZ12ffb} gives the following effective Lagrangian for a quark $f=u,d$:
\begin{align}
\label{eqLeffu}
\mathcal{L}^f_{\mathrm{eff}} &=\fr{g^2}{4c^2_WM^2_{Z_1}}  (\bar{e}\ga_\mu\ga^5e)
\left( \bar{f}\ga^\mu f\right) \left( g^{(1)}_A(e)g^{(1)}_V(f) + g^{(2)}_A(e)g^{(2)}_V(f)
\fr{M^2_{Z_1}}{M^2_{Z_2}}\right) \crn
&=+\fr{G_F}{\sqrt{2}} (\bar{e}\ga_\mu\ga^5e)  \left( \bar{f}\ga^\mu f\right)\times
2\rho \left( g^{(1)}_A(e)g^{(1)}_V(f) + g^{(2)}_A(e)g^{(2)}_V(f) \fr{M^2_{Z_1}}{M^2_{Z_2}}\right)%
\crn&\equiv - \fr{G_F}{2\sqrt{2}} (\bar{e}\ga_\mu\ga^5e)  \left( \bar{f}\ga^\mu f\right)\times
C^{BSM}_1(f),
\end{align}
where  we have denoted
\be\label{eqC1ud}
C^{\mathrm{BSM}}_1(f)\equiv -4 \rho \left( g^{(1)}_A(e)g^{(1)}_V(f) + g^{(2)}_A(e)g^{(2)}_V(f)
\fr{M^2_{Z_1}}{M^2_{Z_2}}\right).
\ee
The parameter $\rho$ is defined  in Eq.~\eq{eq261}.
Then  a nuclear atom
	$^A_ZX$ with $Z$ protons and $N=A-Z$ neutrons
	consisting of $(2Z+ N)$ quarks $u$ and $Z+2N$ quark $d$ in the first family has
	a weak charge determined as follows \cite{Diener:2011jt}
\be \label{eqQWSM}
Q^{\mathrm{BSM}}_W(^A_ZX)= \left[(2Z+N)C^{\mathrm{BSM}}_1(u) + (Z +2N)C^{\mathrm{BSM}}_1(d)\right],
\ee	
In the SM, it has  only neutral boson $Z\equiv Z_1$ with mass $M_Z\equiv M_{Z_1}$,  while  $\fr{M^2_{Z_1}}{M^2_{Z_2}}=0$, $g^{(1)}_{V,A}(f)=g_{V,A}(f)$ with $f=e,u,d$. It can be derived that $\rho=1$ and $ C^{\mathrm{SM}}_1(f) \equiv  - 4  \, g_A(e) g_V(f)$,  resulting to the popular value APV of $^{133}_{55}Cs$ used to compare with experiments, namely
\be
Q^{\mathrm{SM}}_W(^{133}_{55}Cs) = -73.8684.
\label{eq151}
\ee
The latest value of $Q^{\mathrm{SM}}_W(^{133}_{55}Cs)$ including other loop contributions is given in Ref.~\cite{Tanabashi:2018oca}.

From the $Z-Z'$ mixing matrix $O$ given in \eq{eqmixingZZ'}, the states $Z$ and $Z'$ are written as functions of $Z_{1,2}$.
Inserting them into \eq{eqZZpffb} then identifying the two Lagrangians \eq{eqZZpffb} and \eq{eqZ12ffb}, we obtain:
\begin{align}
\label{eqav12}
g^{(1)}_A(f)&= c_\phi  g_A(f)  -s_\phi  g'_A(f)  , \quad g^{(1)}_V(f)= c_\phi  g_V(f)  -s_\phi  g'_V(f) , \crn
g^{(2)}_A(f)&= s_\phi  g_A(f)  +c_\phi  g'_A(f),\quad g^{(2)}_V(f)=s_\phi  g_V(f ) +c_\phi   g'_V(f).
\end{align}
Now, $C^{\mathrm{BSM}}_1(f)$ is determined as follows:
\begin{align}
C^{\mathrm{BSM}}_1(f)&= - 4\rho \left[\left(c^2_\phi  +s^2_\phi \fr{M^2_{Z_1}}{M^2_{Z_2}}
\right) g_A(e)g_V(f)\right.\label{eqC1ud_av}\\
&- \left[g_A(e) g'_V(f) +g'_A(e) g_V(f)\right] \left(1- \fr{M^2_{Z_1}}{M^2_{Z_2}}\right)
s_\phi  c_\phi
\left.+ \left(s^2_\phi  +c^2_\phi  \fr{M^2_{Z_1}}{M^2_{Z_2}}\right)g'_A(e)g'_V(f)  \right]\, .\nn
\end{align}

To keep the approximation up to order of $O\left(\fr{M^2_{Z_1}}{M^2_{Z_2}}\right),$ we take
$c_\phi \simeq1, \; s^2_\phi \simeq0$ in the first term of expression in \eq{eqC1ud_av}
 because $s_\phi \sim O\left(\fr{M^2_{Z_1}}{M^2_{Z_2}}\right)$. Hence, $g^{(1)}_A(e) g^{(1)}_V(f)
 \simeq g_A(e) g_V(f) -[g_A(e) g'_V(f) +g'_A(e) g_V(f)]s_\phi $.  In contrast, the second term
 of \eq{eqC1ud} is simple, $g^{(2)}_A(e)g^{(2)}_V(f)\simeq g'_A(e)g'_V(f)$.

Thus
\begin{align}
\label{eqC1ud_avt}
C^{\mathrm{BSM}}_1(f)&= - 4 \rho \left[g_A(e)g_V(f) - \left[g_A(e) g'_V(f)
+g'_A(e) g_V(f)\right] s_\phi + \left(\fr{M^2_{Z_1}}{M^2_{Z_2}}\right)g'_A(e)g'_V(f) \right] \crn
&+{\cal O}\left(\fr{M^4_{Z_1}}{M^4_{Z_2}}\right)\, .
\end{align}

Let us  now deal with  a  derivation of  the weak charge
\bea
\De Q^{\mathrm{BSM}}_W(^A_ZX) &=&  Q^{\mathrm{BSM}}_W(^A_ZX)  - Q^{\mathrm{SM}}_W(^A_ZX)  \crn
&&
= - 4\left\{ \left(\fr{N-Z}{4}  + Z s_W^2\right) \rho  - \fr{N-Z}{4}  +  Z s_W^2 \right.\crn
&&
- s_\phi \left((2Z+N) \left[g_A(e) g'_V(u) +g'_A(e) g_V(u)\right] \right.
\crn
&&+\left. (Z+2N) \left[g_A(e) g'_V(d) +g'_A(e) g_V(d)\right]  \right)
\crn&&\left.+ \left(\fr{M_{Z_1}^2}{M^2_{Z_2}}\right)[(2Z+N)g'_A(e)g'_V(u) +(Z+2N) g'_A(e)g'_V(d) ]
\right\}
\crn&&+{\cal O}\left(\fr{M^4_{Z_1}}{M^4_{Z_2}}\right),
\eea
where we have used the SM couplings of the electron, quarks $u$ and $d$  given in Table \ref{tab1}
and $\rho  s_\phi \simeq s_\phi \, ,
\rho \left(\fr{M_{Z_1}^2}{M^2_{Z_2}}\right) \simeq \left(\fr{M_{Z_1}^2}{M^2_{Z_2}}\right) $.

To continue, we check the shift of $\de(s_W^2)$ introduced in Ref.~\cite{Altarelli:1991ci}. Using the formula
\be\label{eqsw}
s_W^2c^2_W=\fr{\mu^2}{\rho M_Z^2},\quad \mu\equiv \fr{\pi\alpha}{\sqrt{2} G_F},
\ee
where $\mu$ and $M_Z$ are fixed as experimental inputs.  Defining  $x=s_W^2$, with $c^2_W=1-x$,
as a variable in the following intermediate steps, ones have
\begin{align}\label{eq_shiftsw}
(x-x^2)\rho&=\mathrm{const} \rightarrow 0=\fr{\de }{\de \, x}[(x-x^2)\rho]=(1-2x)\rho
+(x-x^2)\fr{\de \rho}{\de \,x}\crn
\rightarrow& \hs  \de (s_W^2)=\de \,x=-\fr{x-x^2}{(1-2x)\rho}\de  \rho \simeq
-\fr{s_W^2c_W^2}{c_{2W}}\De  \rho \, .
\end{align}
Here we have used that fact that $\rho=1 +\De \rho$ with $\De \rho= \mathcal{O}\left(\fr{M^2_{Z_1}}{M^2_{Z_2}}\right)$.
The result in Eq.~\eq{eq_shiftsw} is consistent with Eq. (2.13) of Ref.~\cite{Altarelli:1991ci},
but slight different from the expression used in Refs.~\cite{Altarelli:1990dt,Altarelli:1996pr,Buras:2014yna}.

To compare with the SM, we have to derive the deviation of $s^2_W$ and $\rho$ from the ones of the SM,
namely $\rho  \, \rightarrow\,  1+\De \rho$ and $s^2_W \rightarrow \, s^2_W +\de (s^2_W)$,
where $\de\, (s^2_W)$ is given in \eq{eq_shiftsw}.

Applying  the above procedure, we have
\bea
&&\De Q^{\mathrm{BSM}}_W(^A_ZX)
= (Z-N) (1+\De \rho) - 4Z[ s_W^2  (1+\De \rho) -\fr{s_W^2 c_W^2}{c_{2W}}\De \rho]  - Z-N  - 4 Z s_W^2 \crn
&&
+ 4 s_\phi \left\{(2Z+N) \left[g_A(e) g'_V(u) +g'_A(e) g_V(u)\right] \right.
\crn
&&+\left. (Z+2N) \left[g_A(e) g'_V(d) +g'_A(e) g_V(d)\right]  \right\}
\crn&&- 4\left(\fr{M_{Z_1}^2}{M^2_{Z_2}}\right)[(2Z+N)g'_A(e)g'_V(u) +(Z+2N) g'_A(e)g'_V(d) ]
+{\cal O}\left(\fr{M^4_{Z_1}}{M^4_{Z_2}}\right).
\label{eq135}
\eea

Substituting $N=A-Z$ into \eq{eq135}, we obtain the expression \eq{eq_DeQBSM} for $\De Q^{\mathrm{BSM}}_W(^A_ZX)$. If the
scale dependence of gauge couplings  are taken into account, replacements need to be done in Eq. \eq{eqVff}, namely
 $g\rightarrow g({M_{Z_{1,2}}})$  for couplings of $Z_{1,2}$, respectively. In addition, the factor in front
 of Eq.~\eq{eqLeffu} is always $g^2(M_{Z_1})$, corresponding to the $M_{Z_1}$ scale.  Hence, the $Z'$ couplings
 in \eq{eq135} should be replaced with $g'_{A,V}(f)\rightarrow g'_{A,V}(f)\times \fr{g(M_{Z_2})}{g(M_{Z_1})}$, resulting in  Eq~\eq{del4}.

To conclude this section, we note that the above procedure can be easily extended for  the Two Higgs Doublet Models with the addition of an Abelian gauge group ~\cite{Campos:2017dgc}, and  the models with
 two or more extra gauge bosons, for instance
the  models based on the gauge group $\mbox{SU}(3)_C\times \mbox{SU}(4)_L \times \mbox{U}(1)_X$
\cite{Pisano:1994tf,Long:2016lmj}.
In the framework of the 3-4-1 model, the APV has been  considered in Ref.~\cite{Nisperuza:2009xm}.

\section{\label{APV331}  General discussions  on recent 3-3-1 models}

The APV can be considered in a more general class of 331 models with arbitrary parameter $\beta$ defined
the electric charge of the model in Eq.~\eq{qoperator}. We consider here the popular class of 3-3-1 models
with three Higgs triplets, namely the 3-3-1\, $\beta$,  where general analytic  ingredient for  determining APV such as  the $Z-Z'$ mixing
$s_\phi$ and heavy neutral gauge boson are well-known \cite{CarcamoHernandez:2005ka,  Buras:2014yna}. Furthermore,
 the formula of APV for  these models was mentioned \cite{CarcamoHernandez:2005ka,Martinez:2014lta}, but it needs to
  be improved, at least   because of  the mixing angle and the scale dependence of the gauge couplings concerned in
  Ref.~\cite{Buras:2014yna}. In addition, many new models with $\beta\neq \pm\fr{1}{\sqrt{3}},
  \pm\sqrt{3}$ such as $\beta=0,\pm\fr{2}{\sqrt{3}}$ discussed recently  should be paid attention to
   \cite{Buras:2014yna,Buras:2012dp,Hue:2015mna}.  The APV  relating with these models will be
    discussed in the following.

Three Higgs triplets   are defined the same as those given in Table 3 of  Ref. \cite{CarcamoHernandez:2005ka}, except that the
VEVs of neutral components are denoted as those in Ref.~\cite{Buras:2014yna} for consistence with
 the definition of $t_v$ appearing in Eq.~\eq{eq_bbeta}. The standard definitions of covariant derivatives were given in
Ref.~\cite{Buras:2012dp}, which are consistent
 with Eq.~\eq{eq_PNC} and
\be\label{eq_t}
t\equiv \fr{g_X}{g}=\fr{\sqrt{6}s_W}{\sqrt{1-(1+\beta^2) s_W^2}}.
\ee
The masses of the SM gauge  bosons including $W^\pm_\mu=\fr{W^1_\mu
 \mp iW^2_\mu}{\sqrt{2}}$ and $Z_\mu$  are
\be\label{eq_WZmass}
M^2_W=\fr{g^2(v^2_\rho +v^2_\eta)}{4},  \quad M^2_Z=\fr{M_W^2}{c_W^2}.
\ee

After the breaking $SU(3)_L\otimes U(1)_X\rightarrow U(1)_Q$, the model consists of
three neutral gauge bosons including one massless photon, a SM  boson $Z_\mu$ and a new
 heavy $Z'_\mu$  \cite{CarcamoHernandez:2005ka}
\begin{align} \label{eq_AZZp}
A_\mu&= s_W W^3_\mu +c_W\left( \beta t_W W^8_\mu +\sqrt{1-\beta^2 t^2_W } B_\mu\right),
\crn~Z_\mu&= c_W W^3_\mu -s_W\left( \beta t_W W^8_\mu +\sqrt{1-\beta^2 t^2_W } B_\mu\right),
\crn~Z'_\mu&= \sqrt{1-\beta^2 t^2_W }  W^8_\mu  - \beta\, t_WB_\mu,
\end{align}
where the state $Z'_\mu$ has an opposite sign with the choice in Ref.~\cite{CarcamoHernandez:2005ka,Buras:2014yna,Martinez:2014lta}
in order to be consistent with the particular case of the 3-3-1 CKS model  we mentioned above.
In the limit $v_\chi \ll v_\rho,v_\eta$, the $Z-Z'$  mixing angle in Eq.~\eq{eq1817}
can be found as  given in Eq.~\eq{eq-sphibe}. We emphasize that this formula  was  introduced firstly in Ref.~\cite{Buras:2014yna},
which corrects the one in Ref.~\cite{CarcamoHernandez:2005ka}.

 We note that our choice of the mixing matrix is
\begin{align}\label{eq_CZZ'}
C_{ZZ'}\equiv \begin{pmatrix}
c_\phi& -s_\phi \\
s_\phi& c_\phi
\end{pmatrix},
\end{align}
which defines the relation between two base of neutral gauge boson states:
$(Z_1,Z_2)^T=C_{ZZ'}(Z,Z')^T$.
The mixing angle $\phi$ in this definition is different from that in Refs. \cite{CarcamoHernandez:2005ka,Buras:2014yna,Martinez:2014lta} by a minus sign.
Combining with the state $Z'$ defined in this work, the formula \eq{eq-sphibe} determining $\phi$ was found to be consistent with Ref.~\cite{Buras:2014yna}.
Based on this, the needed couplings can be calculated, as  given in Table~\ref{table_gV}, where our notations coincide with those in
Ref.~\cite{CarcamoHernandez:2005ka}.   We can see that the mixing
angle $\phi$ and couplings are
consistent with the particular case of $\beta=0$ and $v_\rho=0$ we discussed above.

Now comparing with the result in table 4 of  Ref.~\cite{CarcamoHernandez:2005ka}, we found an global opposite sign of $Z'$ couplings, which can be removed by choosing the  $Z'$ state  to have the same sign defined in
Ref.~\cite{CarcamoHernandez:2005ka}. But   a minus sign will also appear in the right-handed side of Eq.~\eq{eq-sphibe}.
In conclusion, both signs of $s_\phi$ and $Z'$ couplings will be changed if  the phase of  the state $Z'$ is changed,
leading to the fact that the Eq.~\eq{eq_DeQBSM} is independent with the phase of  $Z'$.

Now we will pay attention to the $^{133}_{55}Cs$, where $\left(A- 2.39782 \times Z  \right)
\De \rho\simeq 1.12 \De\rho=\mathcal{O}(10^{-4})\ll |\De Q(Cs)| $ following recent experimental results.
Hence,  in the framework of the 3-3-1\, $\beta$ model,  the expression for APV of  $Cs$  is written as Eq.~\eq{del8}, based on Eq.~\eq{del4}, where the term depending on the $\rho$ parameter can be ignored. For $s_\phi$ given in Eq.~\eq{eq-sphibe}, the respective  $Z'$ couplings are listed in
Table~\ref{table_gV}.


\begin{thebibliography}{99}


\bibitem{Valle:1983dk}
J.~W.~F.~Valle and M.~Singer,
Phys.\ Rev.\ D {\bf 28} (1983) 540.


\bibitem{Pisano:1991ee}
  F.~Pisano and V.~Pleitez, \emph{
  Phys.\ Rev.}\ D {\bf 46}, 410 (1992)
  [hep-ph/9206242].



\bibitem{Foot:1992rh}
  R.~Foot, O.~F.~Hernandez, F.~Pisano and V.~Pleitez, \emph{
  Phys.\ Rev}.\ D {\bf 47}, 4158 (1993)
  [hep-ph/9207264].



\bibitem{Frampton:1992wt}
  P.~H.~Frampton, \emph{
  Phys.\ Rev.\ Lett.}\  {\bf 69}, 2889 (1992).



\bibitem{Hoang:1996gi}
  H.~N.~Long,\emph{
  Phys.\ Rev.}\ D {\bf 54}, 4691 (1996)
  [hep-ph/9607439].



\bibitem{Hoang:1995vq}
  H.~N.~Long, \emph{
  Phys.\ Rev.}\ D {\bf 53}, 437 (1996)
  [hep-ph/9504274].



\bibitem{Foot:1994ym}
  R.~Foot, H.~N.~Long and T.~A.~Tran, \emph{
  Phys.\ Rev.}\ D {\bf 50}, no. 1, R34 (1994)
  [hep-ph/9402243].


\bibitem{deSousaPires:1998jc}
  C.~A.~de Sousa Pires and O.~P.~Ravinez, \emph{
  Phys.\ Rev.}\ D {\bf 58}, 035008 (1998)
  [Phys.\ Rev.\ D {\bf 58}, 35008 (1998)]
  [hep-ph/9803409].



\bibitem{VanDong:2005ux}
  P.~V.~Dong and H.~N.~Long, \emph{
  Int.\ J.\ Mod.\ Phys.}\ A {\bf 21}, 6677 (2006)
  [hep-ph/0507155].



\bibitem{Montero:1998yw}
  J.~C.~Montero, V.~Pleitez and O.~Ravinez, \emph{
  Phys.\ Rev}.\ D {\bf 60}, 076003 (1999)
  [hep-ph/9811280].



\bibitem{Montero:2005yb}
  J.~C.~Montero, C.~C.~Nishi, V.~Pleitez, O.~Ravinez and M.~C.~Rodriguez, \emph{
  Phys.\ Rev.}\ D {\bf 73}, 016003 (2006)
  [hep-ph/0511100].



\bibitem{Pal:1994ba}
  P.~B.~Pal, \emph{
  Phys.\ Rev.}\ D {\bf 52}, 1659 (1995)
  [hep-ph/9411406].


\bibitem{Dias:2002gg}
  A.~G.~Dias, V.~Pleitez and M.~D.~Tonasse, \emph{
  Phys.\ Rev.}\ D {\bf 67}, 095008 (2003)
  [hep-ph/0211107].




\bibitem{Dias:2003zt}
  A.~G.~Dias and V.~Pleitez, \emph{
  Phys.\ Rev}.\ D {\bf 69}, 077702 (2004)
  [hep-ph/0308037].



\bibitem{Dias:2003iq}
  A.~G.~Dias, C.~A.~de S. Pires and P.~S.~Rodrigues da Silva, \emph{
  Phys.\ Rev.}\ D {\bf 68}, 115009 (2003)
  [hep-ph/0309058].



\bibitem{CarcamoHernandez:2017cwi}
A.~E.~C\'arcamo Hern\'andez, S.~Kovalenko, H.~N.~Long and I.~Schmidt, \emph{
JHEP} {\bf 1807} (2018) 144
[arXiv:1705.09169 [hep-ph]].


\bibitem{Froggatt:1978nt}
C.~D.~Froggatt and H.~B.~Nielsen,
\emph{
Nucl.\ Phys.} \ B {\bf 147} (1979) 277.


\bibitem{CarcamoHernandez:2016pdu}
A.~E.~C\'arcamo Hern\'andez, S.~Kovalenko and I.~Schmidt, \emph{
JHEP} {\bf 1702} (2017) 125
[arXiv:1611.09797 [hep-ph]].



\bibitem{Huitu:2017ukq}
  K.~Huitu and N.~Koivunen, \emph{
  Phys.\ Rev.}\ D {\bf 98} (2018) no.1,  011701
  [arXiv:1706.09463 [hep-ph]].


\bibitem{Long:2018dun}
  H.~N.~Long, N.~V.~Hop, L.~T.~Hue, N.~H.~Thao and A.~E.~ C\'arcamo
  Hern\'andez:\emph{ Higgs and gauge boson phenomenology of the 3-3-1 model with CKS mechanism},
  arXiv:1810.00605 [hep-ph].


\bibitem{Freitas:2018vnt}
F.~F.~Freitas, C.~A.~de S. Pires and P.~Vasconcelos,
Phys.\ Rev.\ D {\bf 98} (2018) no.3,  035005
[arXiv:1805.09082 [hep-ph]].

\crb{\bibitem{Aaboud:2017buh}
M.~Aaboud {\it et al.} [ATLAS Collaboration],
JHEP {\bf 1710} (2017) 182
[arXiv:1707.02424 [hep-ex]].

\bibitem{Aaboud:2017sjh}
M.~Aaboud {\it et al.} [ATLAS Collaboration],
JHEP {\bf 1801} (2018) 055
[arXiv:1709.07242 [hep-ex]].

\bibitem{Sirunyan:2018exx}
A.~M.~Sirunyan {\it et al.} [CMS Collaboration],
JHEP {\bf 1806} (2018) 120
[arXiv:1803.06292 [hep-ex]].

\bibitem{Aad:2019fac}
G.~Aad {\it et al.} [ATLAS Collaboration],
``Search for high-mass dilepton resonances using 139 fb$^{-1}$ of $pp$ collision data collected at $\sqrt{s}=$13 TeV with the ATLAS detector,''
arXiv:1903.06248 [hep-ex].

\bibitem{Coriano:2018coq}
G.~Corcella, C.~Corianò, A.~Costantini and P.~H.~Frampton,
Phys.\ Lett.\ B {\bf 785} (2018) 73
[arXiv:1806.04536 [hep-ph]].
}
 \bibitem{Bennett:1999pd}
 S.~C.~Bennett and C.~E.~Wieman,
 Phys.\ Rev.\ Lett.\  {\bf 82} (1999) 2484
 Erratum: [Phys.\ Rev.\ Lett.\  {\bf 82} (1999) 4153]
 Erratum: [Phys.\ Rev.\ Lett.\  {\bf 83} (1999) 889]
 [hep-ex/9903022].
 
 \bibitem{Bouchiat:2004sp}
 C.~Bouchiat and P.~Fayet,
 Phys.\ Lett.\ B {\bf 608} (2005) 87
 [hep-ph/0410260].
 
 
 \bibitem{Davoudiasl:2012qa}
 H.~Davoudiasl, H.~S.~Lee and W.~J.~Marciano,
 Phys.\ Rev.\ Lett.\  {\bf 109} (2012) 031802
 [arXiv:1205.2709 [hep-ph]].
 
 \bibitem{Rosner:2001ck}
 J.~L.~Rosner,
 Phys.\ Rev.\ D {\bf 65} (2002) 073026
 [hep-ph/0109239].
 
 
 \bibitem{Ginges:2003qt}
 J.~S.~M.~Ginges and V.~V.~Flambaum,
 Phys.\ Rept.\  {\bf 397} (2004) 63
 [physics/0309054].
 
 
  \bibitem{Guena:2005uj}
 J.~Guena, M.~Lintz and M.~A.~Bouchiat,
 Mod.\ Phys.\ Lett.\ A {\bf 20} (2005) 375
 [physics/0503143].
 
 

  \bibitem{Erler:2014fqa}
  J.~Erler, C.~J.~Horowitz, S.~Mantry and P.~A.~Souder,
  Ann.\ Rev.\ Nucl.\ Part.\ Sci.\  {\bf 64} (2014) 269
  [arXiv:1401.6199 [hep-ph]].
  
 
  
  
\bibitem{Dzuba:2012kx}
V.~A.~Dzuba, J.~C.~Berengut, V.~V.~Flambaum and B.~Roberts,
Phys.\ Rev.\ Lett.\  {\bf 109} (2012) 203003
[arXiv:1207.5864 [hep-ph]].


\bibitem{Tanabashi:2018oca}
M.~Tanabashi {\it et al.} [Particle Data Group],
Phys.\ Rev.\ D {\bf 98} (2018) no.3,  030001.

\bibitem{Erler:2013xha}
J.~Erler and S.~Su,
Prog.\ Part.\ Nucl.\ Phys.\  {\bf 71} (2013) 119
[arXiv:1303.5522 [hep-ph]].

\bibitem{Souder:2015mlu}
P.~Souder and K.~D.~Paschke,
Front.\ Phys.\ (Beijing) {\bf 11} (2016) no.1,  111301.
doi:10.1007/s11467-015-0482-0

\bibitem{Androic:2018kni}
D.~Andro\'{\i}c {\it et al.} [Qweak Collaboration],
Nature {\bf 557} (2018) no.7704,  207.

\bibitem{Altarelli:1991ci}
G.~Altarelli, R.~Casalbuoni, S.~De Curtis, N.~Di Bartolomeo, F.~Feruglio and R.~Gatto,
Phys.\ Lett.\ B {\bf 261} (1991) 146.


 \bibitem{Hoang:2000jy}
H.~N.~Long and L.~P.~Trung,
Phys.\ Lett.\ B {\bf 502} (2001) 63
[hep-ph/0010204].


\bibitem{CarcamoHernandez:2005ka}
A.~E.~Carcamo Hernandez, R.~Martinez and F.~Ochoa,
Phys.\ Rev.\ D {\bf 73} (2006) 035007
[hep-ph/0510421].

\bibitem{Gutierrez:2005rq}
D.~A.~Gutierrez, W.~A.~Ponce and L.~A.~Sanchez,
Int.\ J.\ Mod.\ Phys.\ A {\bf 21} (2006) 2217
[hep-ph/0511057].

\bibitem{Dong:2006cn}
P.~V.~Dong, H.~N.~Long and D.~T.~Nhung,
Phys.\ Lett.\ B {\bf 639} (2006) 527
[hep-ph/0604199].

\bibitem{Salazar:2007ym}
J.~C.~Salazar, W.~A.~Ponce and D.~A.~Gutierrez,
Phys.\ Rev.\ D {\bf 75} (2007) 075016
[hep-ph/0703300 [HEP-PH]].

\bibitem{Gauld:2013qja}
R.~Gauld, F.~Goertz and U.~Haisch,
JHEP {\bf 1401} (2014) 069
[arXiv:1310.1082 [hep-ph]].




\bibitem{Martinez:2014lta}
R.~Martinez and F.~Ochoa,
Phys.\ Rev.\ D {\bf 90} (2014) no.1,  015028
[arXiv:1405.4566 [hep-ph]].

\bibitem{Buras:2013dea}
A.~J.~Buras, F.~De Fazio and J.~Girrbach,
JHEP {\bf 1402} (2014) 112
[arXiv:1311.6729 [hep-ph]].

\bibitem{Buras:2014yna}
A.~J.~Buras, F.~De Fazio and J.~Girrbach-Noe,
JHEP {\bf 1408} (2014) 039
[arXiv:1405.3850 [hep-ph]].

\bibitem{Beringer:1900zz}
J.~Beringer {\it et al.} [Particle Data Group],
Phys.\ Rev.\ D {\bf 86} (2012) 010001.


\bibitem{Martinez:2006gb}
R.~Martinez and F.~Ochoa,
Eur.\ Phys.\ J.\ C {\bf 51}, 701 (2007)
[hep-ph/0606173].


\bibitem{Ochoa:2005ih}
F.~Ochoa and R.~Martinez,
``Z-Z' mixing in SU(3)(c) x SU(3)(L) x U(1)(X) models with beta arbitrary'', 
hep-ph/0508082.

\bibitem{Chang:2006aa}
D.~Chang and H.~N.~Long,
Phys.\ Rev.\ D {\bf 73}, 053006 (2006)
[hep-ph/0603098].

\bibitem{Peskin:1990zt}
M.~E.~Peskin and T.~Takeuchi,
Phys.\ Rev.\ Lett.\  {\bf 65} (1990) 964.




\bibitem{Hoang:1999yv} 
H.~N.~Long and T.~Inami,
Phys.\ Rev.\ D {\bf 61}, 075002 (2000)
[hep-ph/9902475].


\crb{\bibitem{Hue:2018dqf}
L.~T.~Hue and L.~D.~Ninh,
Eur.\ Phys.\ J.\ C {\bf 79} (2019) no.3,  221
[arXiv:1812.07225 [hep-ph]].
}
\bibitem{Buras:2012dp}
A.~J.~Buras, F.~De Fazio, J.~Girrbach and M.~V.~Carlucci,
JHEP {\bf 1302} (2013) 023
[arXiv:1211.1237 [hep-ph]].

\bibitem{Branco:2011iw}
G.~C.~Branco, P.~M.~Ferreira, L.~Lavoura, M.~N.~Rebelo, M.~Sher and J.~P.~Silva,
Phys.\ Rept.\  {\bf 516} (2012) 1
[arXiv:1106.0034 [hep-ph]].


\bibitem{Ng:1992st}
D.~Ng,
Phys.\ Rev.\ D {\bf 49} (1994) 4805
[hep-ph/9212284].


\bibitem{Dias:2004dc}
A.~G.~Dias, R.~Martinez and V.~Pleitez,
Eur.\ Phys.\ J.\ C {\bf 39} (2005) 101
[hep-ph/0407141].



\bibitem{Frampton:2002st}
P.~H.~Frampton,
Mod.\ Phys.\ Lett.\ A {\bf 18} (2003) 1377
[hep-ph/0208044].

\bibitem{Buras:2016dxz}
A.~J.~Buras and F.~De Fazio,
JHEP {\bf 1608} (2016) 115
[arXiv:1604.02344 [hep-ph]].

 
 \bibitem{Hue:2015mna}
 L.~T.~Hue and L.~D.~Ninh,
 Mod.\ Phys.\ Lett.\ A {\bf 31} (2016) no.10,  1650062
 [arXiv:1510.00302 [hep-ph]].
 
 \bibitem{Komachenko:1989qn}
 Y.~Y.~Komachenko and M.~Y.~Khlopov,
 Sov.\ J.\ Nucl.\ Phys.\  {\bf 51} (1990) 692
 [Yad.\ Fiz.\  {\bf 51} (1990) 1081].
 
 \bibitem{Boucenna:2016qad}
 S.~M.~Boucenna, A.~Celis, J.~Fuentes-Martin, A.~Vicente and J.~Virto,
 JHEP {\bf 1612} (2016) 059
 
 \bibitem{Hue:2016nya}
 L.~T.~Hue, A.~B.~Arbuzov, N.~T.~K.~Ngan and H.~N.~Long,
 Eur.\ Phys.\ J.\ C {\bf 77} (2017) no.5,  346
 [arXiv:1611.06801 [hep-ph]].
 
 \bibitem{He:2017bft}
 X.~G.~He and G.~Valencia,
 Phys.\ Lett.\ B {\bf 779} (2018) 52
 [arXiv:1711.09525 [hep-ph]].
 
 \bibitem{Richard:2013xfa}
 F.~Richard,
 ``A Z-prime interpretation of Bd$\rightarrow$K*mu+mu- data and consequences for high energy colliders,''
 arXiv:1312.2467 [hep-ph].
 
 
 \bibitem{Salazar:2015gxa}
 C.~Salazar, R.~H.~Benavides, W.~A.~Ponce and E.~Rojas,
 JHEP {\bf 1507} (2015) 096
 [arXiv:1503.03519 [hep-ph]].
 
 \bibitem{Coutinho:2013lta}
 Y.~A.~Coutinho, V.~Salustino Guimarães and A.~A.~Nepomuceno,
 Phys.\ Rev.\ D {\bf 87} (2013) no.11,  115014
 [arXiv:1304.7907 [hep-ph]].
 
 \bibitem{Hue:2017lak}
 L.~T.~Hue, L.~D.~Ninh, T.~T.~Thuc and N.~T.~T.~Dat,
 Eur.\ Phys.\ J.\ C {\bf 78} (2018) no.2,  128
 [arXiv:1708.09723 [hep-ph]].
 
 
\bibitem{Long:1999ij} 
H.~N.~Long and V.~T.~Van,
J.\ Phys.\ G {\bf 25}, 2319 (1999)
[hep-ph/9909302].

\bibitem{Diener:2011jt}
R.~Diener, S.~Godfrey and I.~Turan,
Phys.\ Rev.\ D {\bf 86} (2012) 115017
[arXiv:1111.4566 [hep-ph]].


\bibitem{Altarelli:1990dt}
G.~Altarelli, R.~Casalbuoni, D.~Dominici, F.~Feruglio and R.~Gatto,
Nucl.\ Phys.\ B {\bf 342} (1990) 15.

\bibitem{Altarelli:1996pr}
G.~Altarelli, N.~Di Bartolomeo, F.~Feruglio, R.~Gatto and M.~L.~Mangano,
Phys.\ Lett.\ B {\bf 375} (1996) 292
[hep-ph/9601324].

\bibitem{Campos:2017dgc}
M.~D.~Campos, D.~Cogollo, M.~Lindner, T.~Melo, F.~S.~Queiroz and W.~Rodejohann,
JHEP {\bf 1708} (2017) 092
[arXiv:1705.05388 [hep-ph]].

\bibitem{Pisano:1994tf}
F.~Pisano and V.~Pleitez,
Phys.\ Rev.\ D {\bf 51}, 3865 (1995)
[hep-ph/9401272].


\bibitem{Long:2016lmj}
H.~N.~Long, L.~T.~Hue and D.~V.~Loi,
Phys.\ Rev.\ D {\bf 94}, no. 1, 015007 (2016)
[arXiv:1605.07835 [hep-ph]].

\bibitem{Nisperuza:2009xm}
J.~L.~Nisperuza and L.~A.~Sanchez,
Phys.\ Rev.\ D {\bf 80}, 035003 (2009)
[arXiv:0907.2754 [hep-ph]].








%

















\end{thebibliography}
\end{document}